\documentclass[aps,prd,twocolumn,superscriptaddress,preprintnumbers,nofootinbib]{revtex4-1}

\usepackage[margin=1in]{geometry}
\usepackage[utf8]{inputenc}
\usepackage{tikz}
\usepackage[compat=1.1.0]{tikz-feynman}
\usepackage[T1]{fontenc}
\usepackage{amssymb}
\usepackage{xcolor}
\usepackage{amsmath}
\usepackage{slashed}
\usepackage{braket}
\usepackage[normalem]{ulem}
\usepackage{comment}
\usepackage[caption=false]{subfig}
\usepackage{graphics}
\usepackage{graphicx}
\usepackage{adjustbox}
\usepackage{hyperref}
\hypersetup{colorlinks,linkcolor={blue},citecolor={blue},urlcolor={blue}} 

\begin{document}

\preprint{LA-UR-26-26639}

\title{Axion-Like Particle Search with a Hybrid Cherenkov-Scintillation Detector}
\affiliation{Bartoszek~Engineering,~Aurora,~IL~60506,~USA}
\affiliation{Columbia~University,~New~York,~NY~10027,~USA}
\affiliation{University~of~Edinburgh,~Edinburgh,~United~Kingdom}
\affiliation{Embry$-$Riddle~Aeronautical~University,~Prescott,~AZ~86301,~USA }
\affiliation{University~of~Florida,~Gainesville,~FL~32611,~USA}
\affiliation{Los~Alamos~National~Laboratory,~Los~Alamos,~NM~87545,~USA}
\affiliation{Massachusetts~Institute~of~Technology,~Cambridge,~MA~02139,~USA}
\affiliation{Universidad~Nacional~Aut\'{o}noma~de~M\'{e}xico,~CDMX~04510,~M\'{e}xico}
\affiliation{University~of~New~Mexico,~Albuquerque,~NM~87131,~USA}
\affiliation{New~Mexico~State~University,~Las~Cruces,~NM~88003,~USA}
\affiliation{Texas~A$\&$M~University,~College~Station,~TX~77843,~USA}

\author{A.A.~Aguilar-Arevalo}
\affiliation{Universidad~Nacional~Aut\'{o}noma~de~M\'{e}xico,~CDMX~04510,~M\'{e}xico}
\author{S.~Biedron}
\affiliation{Element~Aero,~San~Leandro,~CA~94577,~USA}
\author{J.~Boissevain}
\affiliation{Bartoszek~Engineering,~Aurora,~IL~60506,~USA}
\author{M.~Borrego}
\affiliation{Los~Alamos~National~Laboratory,~Los~Alamos,~NM~87545,~USA}
\author{L.~Bugel\textsuperscript{\dag}}
\affiliation{Massachusetts~Institute~of~Technology,~Cambridge,~MA~02139,~USA}
\author{M.~Chavez-Estrada}
\affiliation{Universidad~Nacional~Aut\'{o}noma~de~M\'{e}xico,~CDMX~04510,~M\'{e}xico}
\author{J.M.~Conrad}
\affiliation{Massachusetts~Institute~of~Technology,~Cambridge,~MA~02139,~USA}
\author{R.L.~Cooper}
\affiliation{Los~Alamos~National~Laboratory,~Los~Alamos,~NM~87545,~USA}
\affiliation{New~Mexico~State~University,~Las~Cruces,~NM~88003,~USA}
\author{J.R.~Distel}
\affiliation{Los~Alamos~National~Laboratory,~Los~Alamos,~NM~87545,~USA}
\author{J.C.~D’Olivo}
\affiliation{Universidad~Nacional~Aut\'{o}noma~de~M\'{e}xico,~CDMX~04510,~M\'{e}xico}
\author{E.~Dunton}
\affiliation{Los~Alamos~National~Laboratory,~Los~Alamos,~NM~87545,~USA}
\author{B.~Dutta}
\affiliation{Texas~A$\&$M~University,~College~Station,~TX~77843,~USA}
\author{D.E.~Fields}
\affiliation{University~of~New~Mexico,~Albuquerque,~NM~87131,~USA}
\author{M.~Gold}
\affiliation{University~of~New~Mexico,~Albuquerque,~NM~87131,~USA}
\author{E.~Guardincerri}
\affiliation{Los~Alamos~National~Laboratory,~Los~Alamos,~NM~87545,~USA}
\author{E.C.~Huang}
\affiliation{Los~Alamos~National~Laboratory,~Los~Alamos,~NM~87545,~USA}
\author{N.~Kamp}
\affiliation{Massachusetts~Institute~of~Technology,~Cambridge,~MA~02139,~USA}
\author{D.~Kim}
\affiliation{University of South Dakota, Vermillion, SD 57069, USA}
\author{K.~Knickerbocker}
\affiliation{Los~Alamos~National~Laboratory,~Los~Alamos,~NM~87545,~USA}
\author{W.C.~Louis}
\affiliation{Los~Alamos~National~Laboratory,~Los~Alamos,~NM~87545,~USA}
\author{C.F.~Macias-Acevedo}
\affiliation{Universidad~Nacional~Aut\'{o}noma~de~M\'{e}xico,~CDMX~04510,~M\'{e}xico}
\author{R.~Mahapatra}
\affiliation{Texas~A$\&$M~University,~College~Station,~TX~77843,~USA}
\author{J.~Mezzetti}
\affiliation{University~of~Florida,~Gainesville,~FL~32611,~USA}
\author{J.~Mirabal}
\affiliation{Los~Alamos~National~Laboratory,~Los~Alamos,~NM~87545,~USA}
\author{M.J.~Mocko}
\affiliation{Los~Alamos~National~Laboratory,~Los~Alamos,~NM~87545,~USA}
\author{D.A.~Newmark}\email{Contact author: dnewmark@mit.edu}
\affiliation{Massachusetts~Institute~of~Technology,~Cambridge,~MA~02139,~USA}
\author{P.~deNiverville}
\affiliation{Los~Alamos~National~Laboratory,~Los~Alamos,~NM~87545,~USA}
\author{C.~O’Connor}
\affiliation{University~of~Florida,~Gainesville,~FL~32611,~USA}
\author{V.~Pandey}
\affiliation{Fermi National Accelerator Laboratory, Batavia, Illinois 60510, USA}
\author{D.~Poulson}
\affiliation{Los~Alamos~National~Laboratory,~Los~Alamos,~NM~87545,~USA}
\author{H.~Ray}
\affiliation{University~of~Florida,~Gainesville,~FL~32611,~USA}
\author{E.~Renner}
\affiliation{Los~Alamos~National~Laboratory,~Los~Alamos,~NM~87545,~USA}
\author{T.J.~Schaub}
\affiliation{University~of~New~Mexico,~Albuquerque,~NM~87131,~USA}
\author{A.~Schneider}
\affiliation{Texas~A$\&$M~University,~College~Station,~TX~77843,~USA}
\author{M.H.~Shaevitz}
\affiliation{Columbia~University,~New~York,~NY~10027,~USA}
\author{D.~Smith}
\affiliation{Embry$-$Riddle~Aeronautical~University,~Prescott,~AZ~86301,~USA }
\author{W.~Sondheim}
\affiliation{Los~Alamos~National~Laboratory,~Los~Alamos,~NM~87545,~USA}
\author{A.M.~Szelc}
\affiliation{University~of~Edinburgh,~Edinburgh,~United~Kingdom}
\author{C.~Taylor}
\affiliation{Los~Alamos~National~Laboratory,~Los~Alamos,~NM~87545,~USA}
\author{A.~Thompson}
\affiliation{Northwestern~University,~Evanston,~IL~60208,~USA}
\author{W.H.~Thompson}
\affiliation{Los~Alamos~National~Laboratory,~Los~Alamos,~NM~87545,~USA}
\author{R.T.~Thornton}
\affiliation{Los~Alamos~National~Laboratory,~Los~Alamos,~NM~87545,~USA}
\author{M.~Tripathi}
\affiliation{University~of~Florida,~Gainesville,~FL~32611,~USA}
\author{R.~Van~Berg}
\affiliation{Bartoszek~Engineering,~Aurora,~IL~60506,~USA}
\author{R.G.~Van~de~Water}
\affiliation{Los~Alamos~National~Laboratory,~Los~Alamos,~NM~87545,~USA}
\author{J.~Zettlemoyer}
\affiliation{Los~Alamos~National~Laboratory,~Los~Alamos,~NM~87545,~USA}

\begingroup
\renewcommand\thefootnote{\dag}
\footnotetext{Deceased}
\endgroup

\collaboration{The CCM Collaboration}

\date{\today}

\begin{abstract}
This analysis presents the first proof-of-concept search for axion-like particles (ALPs) using a hybrid Cherenkov-scintillation detector at a beam-dump facility. The work is based on the Coherent CAPTAIN-Mills (CCM) experiment, a 10-ton liquid argon light collection detector located at Los Alamos National Laboratory. The CCM200 detector is instrumented with 200 photomultiplier tubes (PMTs), providing approximately 50\% photocathode coverage. To enable optical discrimination, 80\% of the PMTs are coated with a wavelength-shifting material while the remaining 20\% are left uncoated. This configuration, combined with nanosecond-scale timing resolution, provides sensitivity to prompt Cherenkov radiation while maintaining efficient detection of liquid argon scintillation light~\cite{CCM:2025kal,CCM:2025dbq}. This analysis constructs four observables that exploit the Cherenkov emission and event topology expected from ALP-induced electromagnetic interactions. These observables are combined into a likelihood-ratio classifier that provides powerful rejection of steady-state backgrounds. No statistically significant excess above the expected background prediction is observed for $10^{-3}~\mathrm{MeV} < m_a < 10~\mathrm{MeV}$. Nevertheless, the improved background rejection enabled by hybrid Cherenkov-scintillation detection allows this analysis to surpass the sensitivity of the previous CCM120 search~\cite{CCM:2021jmk} despite less exposure and demonstrate the physics potential for hybrid Cherenkov-scintillation detectors. 
\end{abstract}

\maketitle

\section{\label{sec:intro}Introduction}
The axion arises from the spontaneous breaking of the global Peccei-Quinn U(1)$_{\mathrm{PQ}}$ symmetry~\cite{Peccei:1977hh}, proposed to explain the observed conservation of Charge-Parity (CP) symmetry in the strong interaction by dynamically solving the strong-CP problem~\cite{Baker:2006ts,Abel:2020pzs}. The resulting QCD axion~\cite{KSVZ_1,KSVZ_2,DFSZ_1,DFSZ_2} is accompanied by a broader class of axion-like particles (ALPs), which arise in many extensions of the Standard Model. In addition to solving the strong-CP problem, light axions and ALPs are also well-motivated dark matter and dark sector candidates~\cite{Lanfranchi:2020crw}.

Although a broad experimental program is dedicated to searching for very light axions and ALPs (ADMX~\cite{ADMX:2025vom}, CAST~\cite{CAST:2024eil}, IAXO~\cite{IAXO:2019mpb}, HAYSTAC~\cite{HAYSTAC:2024jch,HAYSTAC:2025nrt}, and DMRadio~\cite{Ankel:2026zrv} among others), this work targets higher-mass ALP candidates. Accelerator and reactor based experiments (NA62~\cite{NA62:2025yzs}, NA64~\cite{Dusaev:2020gxi,NA64:2020qwq}, SeaQuest~\cite{Berlin:2018pwi}, DarkQuest~\cite{Blinov:2021say}, NEON~\cite{NEON:2024kwv}, and MINER~\cite{Mirzakhani:2025bqz} among others) offer a complementary avenue for exploring MeV to GeV mass ALP parameter space, which remains comparatively less constrained. ALPs in this mass range could still address the strong-CP problem and provide a window into a more extensive dark sector. 

To search for these rare interactions, optical scintillation detectors are a well-established technology that offer large target masses, high light yield, and excellent energy resolution~\cite{Borexino:2008gab,SNO:2021xpa,JUNO:2015sjr,KamLAND:2002uet}. More recently, hybrid optical detectors capable of simultaneously measuring both prompt Cherenkov emission and delayed scintillation light have emerged as a promising approach for improving event reconstruction and background rejection~\cite{CCM:2025kal,Theia:2019non,Anderson:2022lbb,ANNIE:2023yny}. Because Cherenkov radiation is emitted at a characteristic angle, reconstructing both signals provides additional information for particle identification and topology beyond that available from scintillation light alone. 

The Coherent CAPTAIN-Mills (CCM) experiment at the Lujan Spallation Facility of the Los Alamos Neutron Science Center (LANSCE) employs a liquid argon optical detector designed to leverage scintillation and Cherenkov light for neutrino and beyond the Standard Model (BSM) physics searches~\cite{CCM:2021leg,CCM:2021yzc,CCM:2023itc,Dutta:2024yjp}. A previous search for ALPs was performed with the prototype CCM120 detector~\cite{CCM:2021jmk}, using an exposure of $1.79 \times 10^{21}$ protons on target (POT). This prototype detector was instrumented with 120 PMTs and achieved a timing resolution of $\mathcal{O}(10~\mathrm{ns})$, which limited its ability to separate the prompt Cherenkov signal from the slower scintillation emission. Since this separation requires timing resolution approaching $\mathcal{O}(1~\mathrm{ns})$, the CCM120 analysis relied primarily on timing cuts within the 290~ns beam spill window to suppress steady-state backgrounds. Despite this limitation, the analysis demonstrated the potential of accelerator-based neutrino facilities to probe previously unexplored ALP parameter space.

Since the completion of the CCM120 program, both the detector and data processing techniques have undergone substantial upgrades. The current CCM200 detector features 200 PMTs, an external muon-tagging veto system, and improved data acquisition capabilities enabling more extensive calibration studies. A key feature of the CCM200 design is the combination of wavelength-shifting coated and uncoated PMTs to enable separation of scintillation and Cherenkov light. Specifically, 80\% of the PMTs and the detector walls are evaporatively coated with the wavelength shifter tetraphenyl butadiene (TPB), while the remaining 20\% of PMTs are uncoated. The TPB converts the 128~nm liquid argon scintillation light into visible wavelengths detectable by the PMTs, while the uncoated PMTs retain enhanced sensitivity to the prompt visible component of the broad-spectrum Cherenkov radiation.

In addition to these detector improvements, Ref.~\cite{CCM:2025dbq} presents new calibration and data processing techniques developed for the CCM200 program. These include pulse unfolding to reconstruct single photoelectron hits with 2~ns timing resolution and calibration studies using a $^{22}$Na radioactive source producing $\mathcal{O}(1~\textrm{MeV})$ gamma rays. The combination of wavelength-selective PMTs and improved timing resolution enabled event-by-event separation of scintillation and Cherenkov light from sub-MeV electrons~\cite{CCM:2025kal}. This capability establishes CCM200 as a hybrid optical detector that combines the high light yield of scintillation emission with the directional and particle identification information provided by Cherenkov radiation.

The ability to identify Cherenkov light provides a new handle for distinguishing signal-like electromagnetic interactions from beam-related and environmental backgrounds that are difficult to separate using scintillation information alone. The distinct spatial and temporal signatures of Cherenkov and scintillation light enable the construction of new discriminating observables, which are incorporated in this analysis through the event selection. Building upon the CCM200 data processing and optical model calibration framework established in Ref.~\cite{CCM:2025kal} and Ref.~\cite{CCM:2025dbq}, this analysis improves background rejection and enhances sensitivity to ALP-induced signals compared to the CCM120 result. It provides a proof-of-principle demonstration of hybrid Cherenkov-scintillation detection in a beam-dump experiment and represents the first ALP search performed with a hybrid optical detector at an accelerator-based beam-dump facility. Further details of this analysis are presented in Ref.~\cite{Newmark:2026sea}.

In this work, we revisit the ALP search using data collected during approximately eight weeks in 2022 with the CCM200 detector, corresponding to a total of $1.23 \times 10^{21}$ POT. This analysis considers ALP masses $10^{-3}~\mathrm{MeV} < m_a < 10~\mathrm{MeV}$ and photon coupling strengths $10^{-6}~\mathrm{GeV}^{-1} < g_{a\gamma} <10^{-3}~~\mathrm{GeV}^{-1}$. During data collection, PMT waveforms were recorded in a 16~$\mu$s window around each beam spill, allowing for data-driven background characterization. The data collected before the beam spill are used to characterize the steady-state background, referred to as the ``prebeam'' sample. These events are dominated by low-energy activity from electronics noise, ambient radioactivity in the detector hall, and cosmogenic backgrounds.

In addition to the prebeam sample, this work uses events from the neutron-dominated time region as a data-driven control sample to validate the electromagnetic event selection. These neutron-induced events are generally higher in energy than those in the prebeam sample, providing a complementary test of the selection at higher energies. Following the proton beam interaction with the tungsten target, prompt neutrinos and near-speed-of-light BSM particles (such as ALPs) reach the detector with minimal delay, whereas the larger neutron flux arrives approximately 200~ns later because of their larger mass and the shielding between the target and detector, as described in Ref.~\cite{CCM:2021leg}. This separation in arrival time allows beam timing requirements to isolate the prompt signal region while rejecting neutron-induced backgrounds. Although neutron events are therefore not expected to contribute significantly in the physics region of interest, the neutron-dominated sample provides an important data-driven cross-check of the selection.

\section{\label{sec:signal_model}ALP Signal Model}
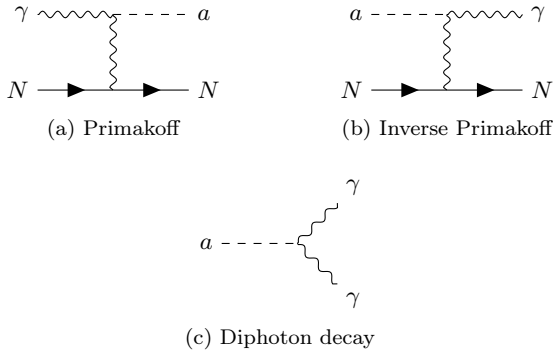
\begin{figure}[h]
\centering
\subfloat[Primakoff]{
     \begin{tikzpicture}
              \begin{feynman}
         \vertex (o1);
         \vertex [right=1cm of o1] (f1) {\(a\)};
         \vertex [left=1cm of o1] (i1){\(\gamma\)} ;
         \vertex [below=1cm of o1] (o2);
         \vertex [right=1cm of o2] (f2) {\(N\)};
         \vertex [left=1cm of o2] (i2) {\(N\)};

         \diagram* {
           (i1) -- [boson] (o1) -- [scalar] (f1),
           (o1) -- [boson] (o2),
           (i2) -- [fermion] (o2),
           (o2) -- [ fermion] (f2),
         };
        \end{feynman} 
       \end{tikzpicture}
       \label{fig:axionPrimakoff}
}\hspace{30pt}
\subfloat[Inverse Primakoff]{  
       \begin{tikzpicture}
              \begin{feynman}
         \vertex (o1);
         \vertex [right=1cm of o1] (f1) {\(\gamma\)};
         \vertex [left=1cm of o1] (i1){\(a\)} ;
         \vertex [below=1cm of o1] (o2);
         \vertex [right=1cm of o2] (f2) {\(N\)};
         \vertex [left=1cm of o2] (i2) {\(N\)};

         \diagram* {
           (i1) -- [scalar] (o1) -- [boson] (f1),
           (o1) -- [boson] (o2),
           (i2) -- [fermion] (o2),
           (o2) -- [fermion] (f2),
         };
        \end{feynman}
       \end{tikzpicture}
       \label{fig:axionInvPrimakoff}
} \\
\subfloat[Diphoton decay]{
       \begin{tikzpicture}
       \begin{feynman}
         \vertex (o1);
         \vertex [left=1cm of o1] (i) {\(a\)};
         \vertex [above right=0.75cm of o1] (f1) {\(\gamma\)};
         \vertex [below right=0.75cm of o1] (f2) {\(\gamma\)};

         \diagram* {
           (i) -- [scalar] (o1),
           (o1) -- [boson] (f1),
           (o1) -- [boson] (f2),
         };
        \end{feynman}
       \end{tikzpicture}
        \label{fig:axionDecayDiphoton}
} \\
    \caption{Diagrams of production and detection processes considered in this analysis. The ALP production mechanism is driven by the Primakoff process in the target while detection is through a combination of both inverse Primakoff scattering and diphoton decay in the detector.}
    \label{fig:axion_feynman}
\end{figure}

This analysis focuses on ALPs that couple predominantly to photons via the effective Lagrangian given in Eq.~(\ref{eq:alp_lagrangian}), where $g_{a\gamma}$ denotes the ALP-photon coupling strength, $a$ represents the ALP field, and $F_{\mu\nu}$ ($\tilde{F}^{\mu\nu}$) is the (dual) field strength tensor~\cite{10.1093/ptep/ptac097}. The production and detection mechanisms considered here are illustrated in Fig.~\ref{fig:axion_feynman}.

\begin{equation}
    \mathcal{L}_{a\gamma\gamma} = -\frac{1}{4} g_{a\gamma} a F_{\mu\nu} \tilde{F}^{\mu\nu}
\label{eq:alp_lagrangian}
\end{equation}

In fixed-target configurations, photon-coupled ALPs are primarily produced via the Primakoff process, in which an incoming photon converts into an ALP through coherent scattering off the nuclear Coulomb field. The corresponding cross section scales approximately as $Z^2$, providing a significant rate enhancement in high-$Z$ target materials such as the tungsten target at the Lujan Spallation Facility.

The inverse Primakoff process, where an ALP converts back into a photon within the nuclear electric field, serves as a complementary detection channel. This mechanism is especially critical for low-mass ALPs ($m_a \lesssim \mathcal{O}(10~\mathrm{keV})$), whose macroscopic decay lengths suppress visible decays within the detector volume. 

For heavier ALPs, the dominant experimental signature is the diphoton decay $a \rightarrow \gamma\gamma$. In this process, the total decay width is given by Eq.~(\ref{eq:diphoton_decay_width}), which depends on the coupling strength $g_{a\gamma}$ and the ALP mass $m_a$. Because the proper lifetime $\tau = 1/\Gamma$ scales as $m_a^{-3}$, heavier ALPs decay rapidly enough to be observed inside the detector. 

\begin{equation}
    \Gamma_{a\rightarrow\gamma\gamma} = \frac{g^2_{a\gamma\gamma} m_a^3}{64 \pi} 
\label{eq:diphoton_decay_width}
\end{equation}

The cross sections for these channels are modeled using the \texttt{alplib} framework~\cite{thompson_alplib_2023}, building on the previous CCM120 search for ALPs~\cite{CCM:2021jmk}. Similar to that work, the Mark-IV target configuration~\cite{markiv_target} is fully described in \texttt{GEANT4}~\cite{GEANT4:2002zbu} to simulate the expected initial gamma-ray flux. For this simulation, particle interactions are modeled with the \texttt{QGSP\_BIC\_AllHP} physics list, which includes comprehensive treatments of electromagnetic processes, low-energy hadronic interactions, and nuclear de-excitation. 

The simulated gamma-ray flux is provided as input to the \texttt{SIREN} injection toolkit~\cite{Schneider:2024eej} to model ALP production, propagation, and detection. The generated final-state photons are subsequently processed through the full detector simulation chain. This includes optical photon production and transport in the \texttt{GEANT4}-based detector optical model (presented in Ref.~\cite{CCM:2025dbq}), PMT response simulation, detector timing effects, and data-driven noise overlays. The signal model is constructed on a grid of $100 \times 100$ values in mass and coupling, corresponding to $10^4$ points in parameter space. The mass range spans $10^{-3}$~MeV to 10~MeV, while the coupling range extends from $10^{-6}$~GeV$^{-1}$ to $10^{-3}$~GeV$^{-1}$; both are sampled uniformly in logarithmic scale. 

\section{\label{sec:event_selection}Event Selection Leveraging Cherenkov Radiation}
The primary goal of this work is to incorporate the characteristic temporal and spatial signatures of electromagnetic interactions from ALPs into the event selection procedure, with particular emphasis on Cherenkov radiation. By exploiting both scintillation and Cherenkov light, this analysis demonstrates the potential of a hybrid optical detector to improve signal discrimination in a high-rate beam dump environment.

The event selection begins with the position and energy reconstruction techniques described in Appendix~\ref{sec:app}. Candidate events must satisfy requirements on reconstructed charge, energy, and fiducial position. Specifically, the reconstructed charge within the first 90 ns must lie between 70~PE and 3000~PE, the reconstructed energy between 0.2~MeV and 10~MeV, the reconstructed radial position within 80~cm of the detector origin, and the reconstructed axial position within $|Z|<40$~cm. The upper energy threshold is set by the current detector calibration, which is based primarily on an MeV-scale $^{22}$Na source~\cite{CCM:2025dbq}. Because the present calibration is limited to MeV-scale data, the reconstructed energy selection is conservatively capped at 10~MeV. Ongoing studies using Michel electrons from cosmic-ray muon decays will extend the validated energy range to approximately 50~MeV.

Although these charge, energy, and fiducial selections substantially reduce the background rate, they do not fully exploit the timing and spatial information contained in the optical signal. Cherenkov photons, despite representing only a small fraction of the detected light, provide complementary information through their prompt arrival times, directional emission, and characteristic spatial distribution.

Applying the Cherenkov-identification techniques developed for calibration data to the full physics dataset is considerably more challenging. For the calibration data, Cherenkov events were identified by requiring hits on uncoated PMTs in the early-time region, which selects the visible component of the Cherenkov emission. Unlike the centrally located sub-MeV $^{22}$Na events studied in Ref.~\cite{CCM:2025kal}, ALP interactions span a wider energy range and occur throughout the detector volume, introducing variations in photon propagation, PMT acceptance, and early-time scintillation backgrounds that reduce the purity of the Cherenkov signal.

Reliable identification of the prompt Cherenkov component first requires an accurate determination of the event start time. This analysis uses a constant fraction discrimination (CFD) algorithm, which defines the event time as the point at which the integrated charge reaches 20\% of its total value. Using this common timing reference, the discriminating observables described below are evaluated for prebeam steady-state backgrounds, simulated ALP events, and the neutron-dominated control sample to quantify their ability to separate electromagnetic signals from background interactions. For many of these variables, a common time window of $-6~\mathrm{ns}$ to $-2~\mathrm{ns}$ relative to the CFD-defined event start time is utilized and will be referred to as the ``prompt'' time window.

Four discriminating observables are constructed to exploit complementary signatures of electromagnetic interactions. Collectively, they probe (i) enhanced prompt-light collection by the uncoated PMTs, (ii) the directional asymmetry of the earliest detected photons, (iii) pulse-shape differences arising from prompt and delayed optical components, and (iv) the overall spatial extent of the reconstructed event. Together, they provide additional discrimination between electromagnetic signal events and background interactions, highlighting the advantages of hybrid optical detection in background rejection. In the rest of this section, these variables are evaluated on the prebeam steady-state background sample, the entire ALP Monte Carlo sample ($10^{-3}~\mathrm{MeV} < m_a < 10~\mathrm{MeV}$), and the neutron-dominated sample. 

\subsection{Number of Hits on the Uncoated PMTs}
The first variable used for event discrimination is the number of reconstructed photoelectron (PE) hits recorded by the uncoated PMTs within the prompt time window. This time interval is selected to maximize sensitivity to the prompt optical signal, where the contribution from Cherenkov photons is expected to be greatest.

\begin{figure}[h]
  \centering
  \includegraphics[width=\linewidth]{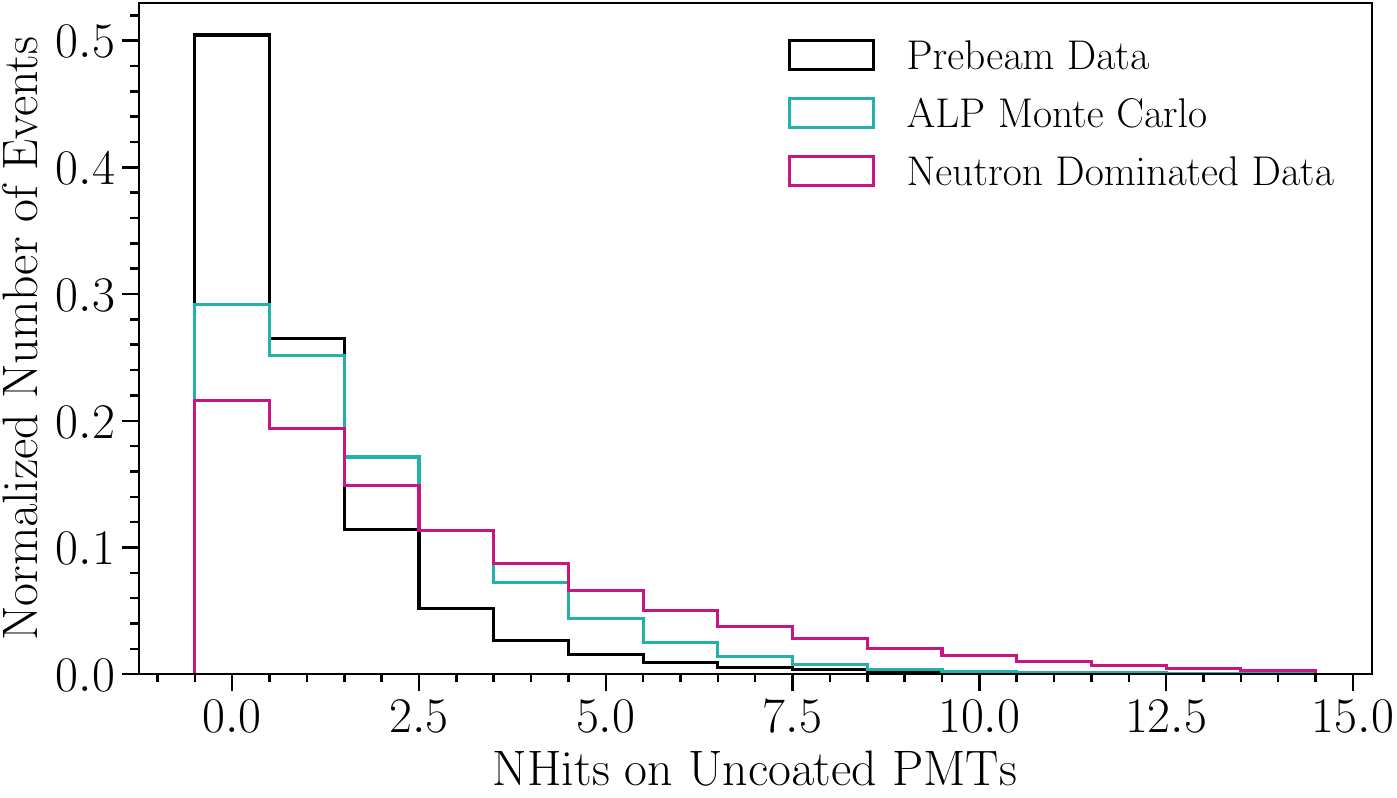}
  \caption{Number of hits on the uncoated PMTs for the prebeam data events, ALP Monte Carlo, and neutron-dominated data events.}
  \label{fig:llr_uncoated_hits}
\end{figure}

The uncoated PMTs are particularly well suited for identifying Cherenkov light during this early-time interval. Because they lack a wavelength-shifting coating, they have little sensitivity to the vacuum ultraviolet scintillation photons and instead primarily detect wavelength-shifted scintillation light together with the visible component of Cherenkov radiation. Consequently, prompt hits on the uncoated PMTs are preferentially associated with visible Cherenkov light, whereas scintillation photons are delayed by both the intrinsic scintillation emission time constant and the additional time required for the photons to propagate to the TPB, undergo wavelength shifting, and reach the PMTs.

Fig.~\ref{fig:llr_uncoated_hits} compares the distributions of the prompt uncoated PMT hit multiplicity for prebeam data, simulated ALP events, and neutron-dominated data. Both the prebeam and ALP samples are markedly peaked at zero hits, indicating that relatively few photons are detected in this narrow prompt window. Nevertheless, the ALP simulation contains a noticeably larger population of events with one or more hits, as expected from the prompt Cherenkov light associated with electromagnetic final states.

The neutron-dominated sample exhibits clearly different behavior, producing a much wider distribution with a substantially larger number of events containing multiple prompt hits on the uncoated PMTs. This trend is consistent with the increased scintillation light generated by higher-energy interactions, which produces greater spillover into the early-time region.

These results demonstrate that the simple hit-count threshold on the uncoated PMTs employed in the $^{22}$Na calibration analysis (the subject of Ref.~\cite{CCM:2025kal}) provides limited discrimination for the ALP search. At higher energies and for interactions occurring throughout the detector volume, the prompt uncoated PMT hit multiplicity alone provides only limited separation between Cherenkov-rich electromagnetic events and scintillation-dominated backgrounds. This limitation motivates combining multiple complementary observables within a multivariate discrimination framework.

\subsection{Spread in the Directionality}
The next variable utilizes the spatial anisotropy of Cherenkov radiation for signal isolation. Unlike scintillation emission which is isotropic, Cherenkov light is directional, offering a robust metric for event discrimination. To calculate this, we restrict the analysis to hits within the prompt time window. Within this interval, we compute the directionality metric $C$, defined in Eq.~(\ref{eq:cid}), which evaluates the angular uniformity of the triggered PMTs.

Specifically, $C$ represents the squared magnitude of the charge-weighted average of unit vectors pointing from the reconstructed interaction vertex ($\vec{x}_{\text{vertex}}$) to each hit PMT position ($\vec{x}_i$) where $q_i$ is the charge of the $i$-th PMT and $Q = \sum_i q_i$ represents the cumulative charge inside the prompt window.

\begin{equation}
    C = \left | \left | \sum_i  \left(\frac{q_i}{Q}\right) \frac{\vec{x}_i - \vec{x}_{\text{vertex}}}{|| \vec{x}_i - \vec{x}_{\text{vertex}} ||} \right | \right |^2
\label{eq:cid}
\end{equation}

Bounded between 0 and 1, $C$ approaches 0 for uniform, isotropic distributions and nears 1 for highly co-linear light emission. Electromagnetic events containing Cherenkov light are therefore expected to exhibit larger values of $C$ than isotropic scintillation-dominated backgrounds.

\begin{figure}[h]
  \centering
  \includegraphics[width=\linewidth]{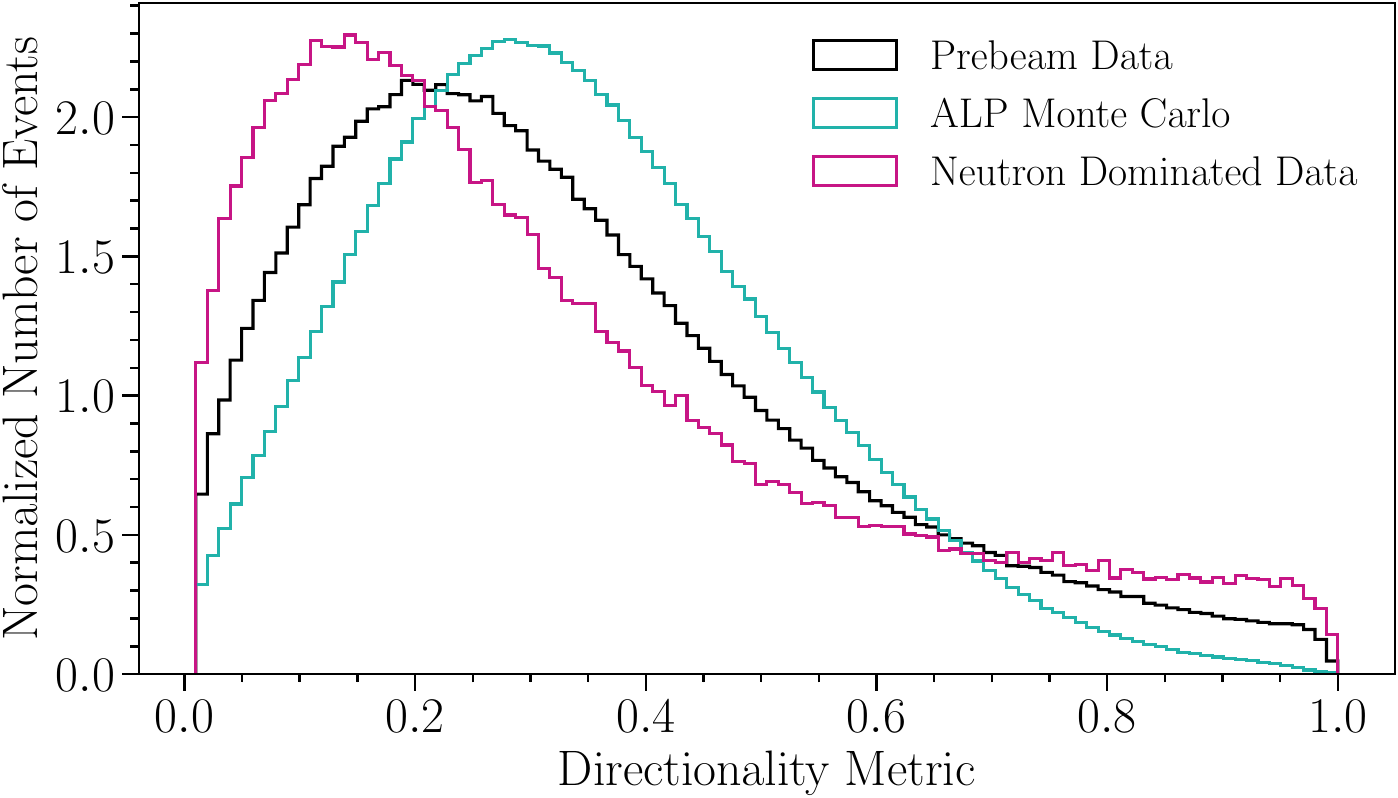}
  \caption{Distribution of the directionality metric $C$, defined in Eq.~(\ref{eq:cid}), across the analyzed datasets. The ALP Monte Carlo events show higher values due to directional photon emission, while neutron-dominated events skew lower, as expected for isotropic light.}
  \label{fig:llr_cid}
\end{figure}

Fig.~\ref{fig:llr_cid} contrasts the $C$ distributions for prebeam data, ALP simulations, and neutron-dominated events. The ALP signal is shifted toward larger values (centered around $C\sim0.3$), reflecting the directional Cherenkov light produced by electromagnetic final states. In contrast, neutron-dominated events cluster at smaller values (around $C\sim0.15$), where the optical signal is primarily composed of isotropic scintillation light. The prebeam sample occupies an intermediate region (around $C\sim0.2$), consistent with a mixture of electromagnetic backgrounds, detector noise, and other background processes. This variable isolates a fundamental physical difference between signal and background, making it a key component of the joint likelihood discriminant.

\subsection{Pulse Shape Ratio}
The third variable included in the likelihood ratio is based on the temporal structure of the detected optical pulse. In traditional liquid argon scintillation detectors, pulse shape discrimination (PSD) is commonly used to distinguish nuclear recoils from electromagnetic interactions by taking advantage of differences in their scintillation decay profiles, which arise from variations in ionization density ($dE/dx$)~\cite{DEAP:2021axq,FIORILLO2006372,Agostini:2015boa,DarkSide-20k:2017zyg}. In the present analysis, however, a conventional PSD technique cannot be implemented because of the operating conditions of the experiment.

The CCM experiment operates in a pulsed beam environment with substantial beam-related activity. Conventional PSD integrates charge over several microseconds to capture the long-lived triplet-state scintillation emission, but such long integration windows would be highly susceptible to pile-up. The charge integration window is therefore limited to $\mathcal{O}(100~\mathrm{ns})$, balancing photo-statistics against pile-up contamination.

Although this integration window precludes conventional PSD, the pulse time evolution still provides useful discriminating information. A pulse shape ratio (PSR) is constructed as the ratio of the charge collected during the early portion of the pulse to the total prompt charge, as defined in Eq.~(\ref{eq:psr}). All timing quantities are measured relative to the CFD-determined event start time.

\begin{equation}
    \text{PSR} = \frac{Q (0~\text{ns} < t < 20~\text{ns})}{Q (0~\text{ns} < t < 90~\text{ns})}
    \label{eq:psr}
\end{equation}

\begin{figure}[h]
    \centering
    \includegraphics[width=\linewidth]{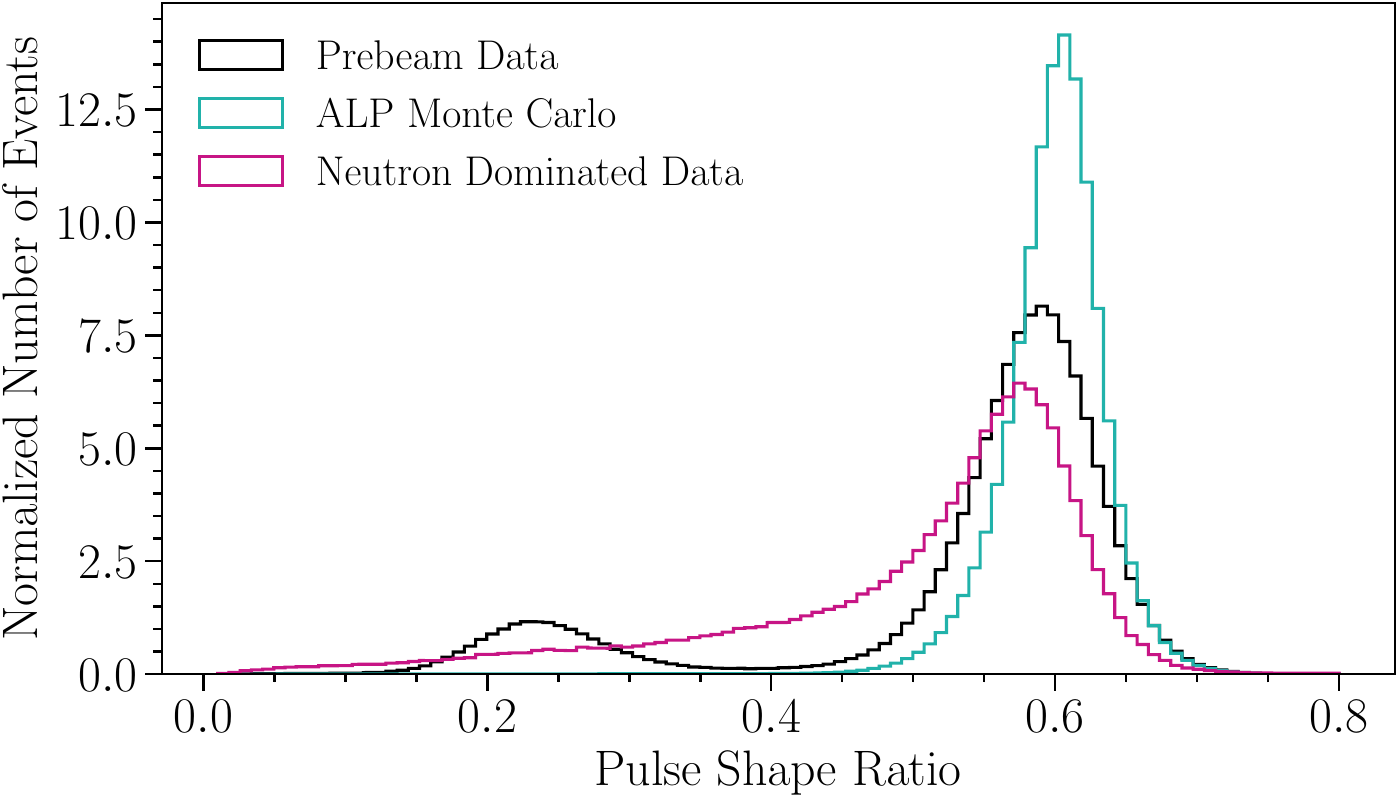}
    \caption{Pulse shape ratio distributions across the ALP, prebeam, and neutron-dominated datasets. While the ALP Monte Carlo events are narrowly peaked around 0.6, the prebeam and neutron-dominated datasets are broader and skewed towards lower values.}
    \label{fig:llr_pulse_shape}
\end{figure}

The PSR distributions for simulated ALP events, prebeam data, and neutron-dominated events are shown in Fig.~\ref{fig:llr_pulse_shape}. The simulated ALP sample exhibits a narrow peak near $\sim0.6$, while the prebeam and neutron-dominated distributions are broader and shifted toward smaller values, consistent with a larger delayed-light contribution. The prebeam data also contain a distinct population near $\sim 0.25$, attributed to electronics noise and overlapping events. Although the PSR is not a conventional PSD variable, it provides complementary timing information that improves signal to background discrimination.

\subsection{Spatial Spread of Events}
The final input to the likelihood ratio is designed to quantify the spatial distribution of the detected charge. This variable takes advantage of the differing event topologies expected for signal and background interactions. ALP-induced events, which produce one or two final-state gamma rays, generally deposit their energy in relatively compact electromagnetic showers. By comparison, neutron backgrounds often undergo multiple scattering and secondary interactions within the detector, resulting in a more diffuse spatial distribution of charge.

\begin{equation}
    \text{RMS} = \sqrt{ \frac{\sum_i q_i ||\vec{x}_i - \vec{x}_{\text{vertex}}||^2}{\sum_i q_i} }
    \label{eq:rms}
\end{equation}

This behavior is quantified using the charge-weighted root-mean-square (RMS) distance of the PMT hits from the reconstructed interaction vertex, as defined in Eq.~(\ref{eq:rms}). Here, $\vec{x}_i$ denotes the position of the $i$-th PMT, $\vec{x}_{\text{vertex}}$ is the reconstructed event vertex, and $q_i$ is the charge measured by that PMT.

The RMS is calculated using only hits within the prompt time window. Restricting the calculation to this interval suppresses contributions from delayed and wavelength-shifted scintillation light, which would otherwise broaden the measured charge distribution. As a result, the RMS provides a measure of the characteristic spatial scale of the prompt light: compact energy depositions produce smaller RMS values, whereas extended or multi-site interactions yield larger values.

\begin{figure}[h]
    \centering
    \includegraphics[width=\linewidth]{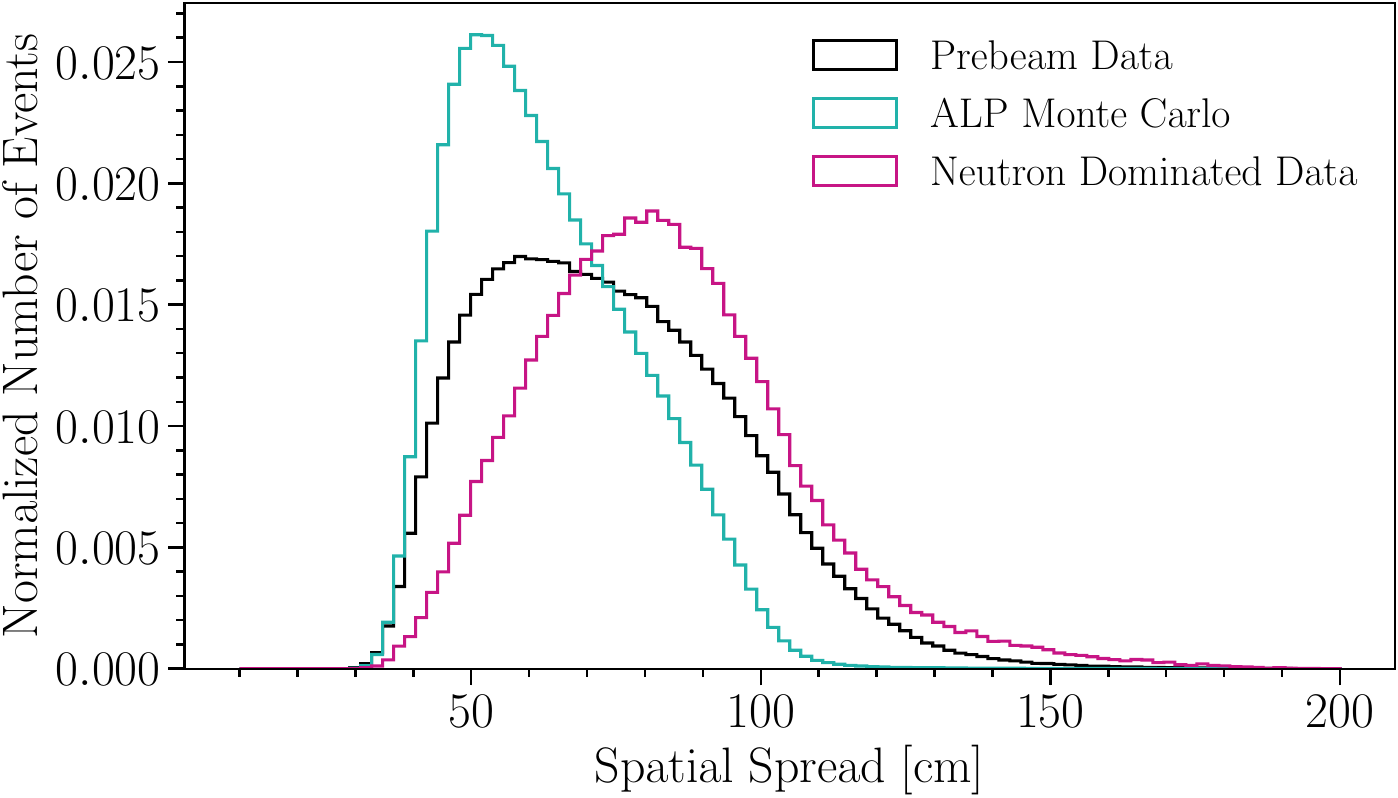}
    \caption{Spatial extent of the early reconstructed charge distribution. The RMS, defined in Eq.~(\ref{eq:rms}), distinguishes localized electromagnetic energy depositions from more spatially extended backgrounds.}
    \label{fig:llr_rms}
\end{figure}

The resulting RMS distributions for the ALP Monte Carlo, prebeam data, and neutron-dominated data are shown in Fig.~\ref{fig:llr_rms}. The simulated ALP events are concentrated at lower RMS values, with a peak near $\sim 50~\mathrm{cm}$, reflecting the compact topology of electromagnetic interactions. In contrast, the neutron-dominated sample peaks around $\sim 80~\mathrm{cm}$ and exhibits a broader distribution, consistent with the larger spatial extent expected from multiple neutron scatters. The prebeam data fall between these extremes, with a maximum near $\sim 60~\mathrm{cm}$ and an extended tail toward larger RMS values, indicative of a mixture of physical backgrounds and non-physical events.

\subsection{Likelihood Ratio Test}
The four observables introduced above are combined into a single discriminant using a log-likelihood ratio (LLR). For each variable, $x_k$, probability density functions (PDFs) are constructed under both the signal and background hypotheses. Signal PDFs are obtained from the simulated ALP sample, while background PDFs are derived from the prebeam data.

\begin{equation}
    \mathrm{LLR} = \sum_{k} \log(p_{\mathrm{sig}}(x_k)) - \log(p_{\mathrm{bkg}}(x_k))
    \label{eq:llr}
\end{equation}

The likelihood for a given event is computed as the product of the individual PDFs for the four observables. The LLR, given by Eq.~(\ref{eq:llr}), is therefore equal to the logarithm of the ratio between the signal and background likelihoods, where $p_{\mathrm{sig}}(x_k)$ and $p_{\mathrm{bkg}}(x_k)$ denote the corresponding signal and background PDFs evaluated at the measured value of the $k$-th observable.

\begin{figure}[h]
  \centering
  \includegraphics[width=\linewidth]{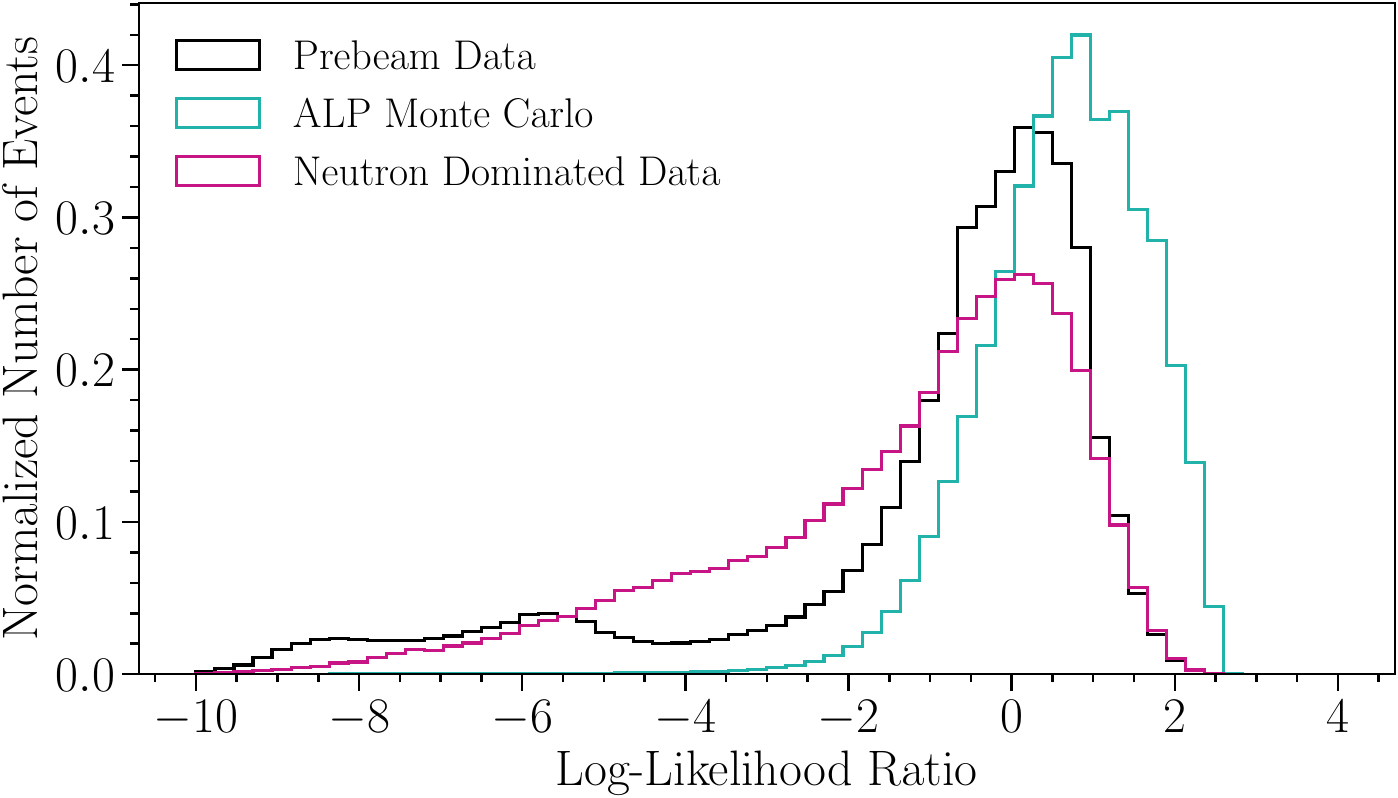}
  \caption{Log-likelihood ratio (LLR) discriminant formed from the four event-level observables described previously. Signal events preferentially occupy larger LLR values than either the prebeam or neutron-dominated background samples. The analysis defines the signal region by requiring $\mathrm{LLR}>1$, balancing signal efficiency against background rejection.}
  \label{fig:final_llr}
\end{figure}

The PDFs are generated from normalized histograms of each variable and interpolated to obtain continuous probability estimates. To prevent numerical instabilities in sparsely populated regions, a small regularization term is included in each PDF.

Fig.~\ref{fig:final_llr} presents the resulting LLR distributions for the ALP simulation, prebeam data, and neutron-dominated data. Signal events preferentially populate larger LLR values, with a broad maximum near 1, whereas both background samples are concentrated at lower values with more pronounced tails. A requirement of LLR $>$ 1 is adopted to define the final signal region, as it provides good background rejection while retaining high signal acceptance.

\subsection{Final Event Selection}
The event selection is optimized to reject prebeam backgrounds while maintaining sensitivity to the low-energy electromagnetic signatures expected from ALP decays. The complete selection and the corresponding efficiencies for the prebeam sample are summarized in Table~\ref{table:alp_cuts}.

\begin{table}[b]
  \caption{Selection efficiencies of cuts utilized in this analysis for the prebeam data sample.}
  \begin{ruledtabular}
    \begin{tabular}{cc}
      Cut &  Efficiency [\%]  \\
      \hline
        Charge & 49.75 \\
        Position  & 18.29 \\
        Energy  & 80.93 \\
        Charge + Position + Energy & 7.19 \\
        Charge + Position + Energy + LLR & 0.49 \\
    \end{tabular}
  \end{ruledtabular}
  \label{table:alp_cuts} 
\end{table}

Basic charge, reconstructed energy, and reconstructed position cuts are applied, as described at the beginning of Sec.~\ref{sec:event_selection}. After those cuts, the final selection criterion is imposed on the log-likelihood ratio described in the previous section. Events with LLR $>$ 1 are retained, providing additional separation between signal-like electromagnetic interactions and residual backgrounds. Together, these requirements reduce the prebeam sample to 0.49\% of its original size. The previous iteration of this analysis using the prototype detector, described in Ref.~\cite{CCM:2021jmk}, achieved 3.2\% efficiency for the prebeam data sample. The detector upgrades allowing use of the Cherenkov signal enhanced the steady state background rejection capabilities by approximately six-fold. 

\begin{figure}[h]
  \centering
  \includegraphics[width=\linewidth]{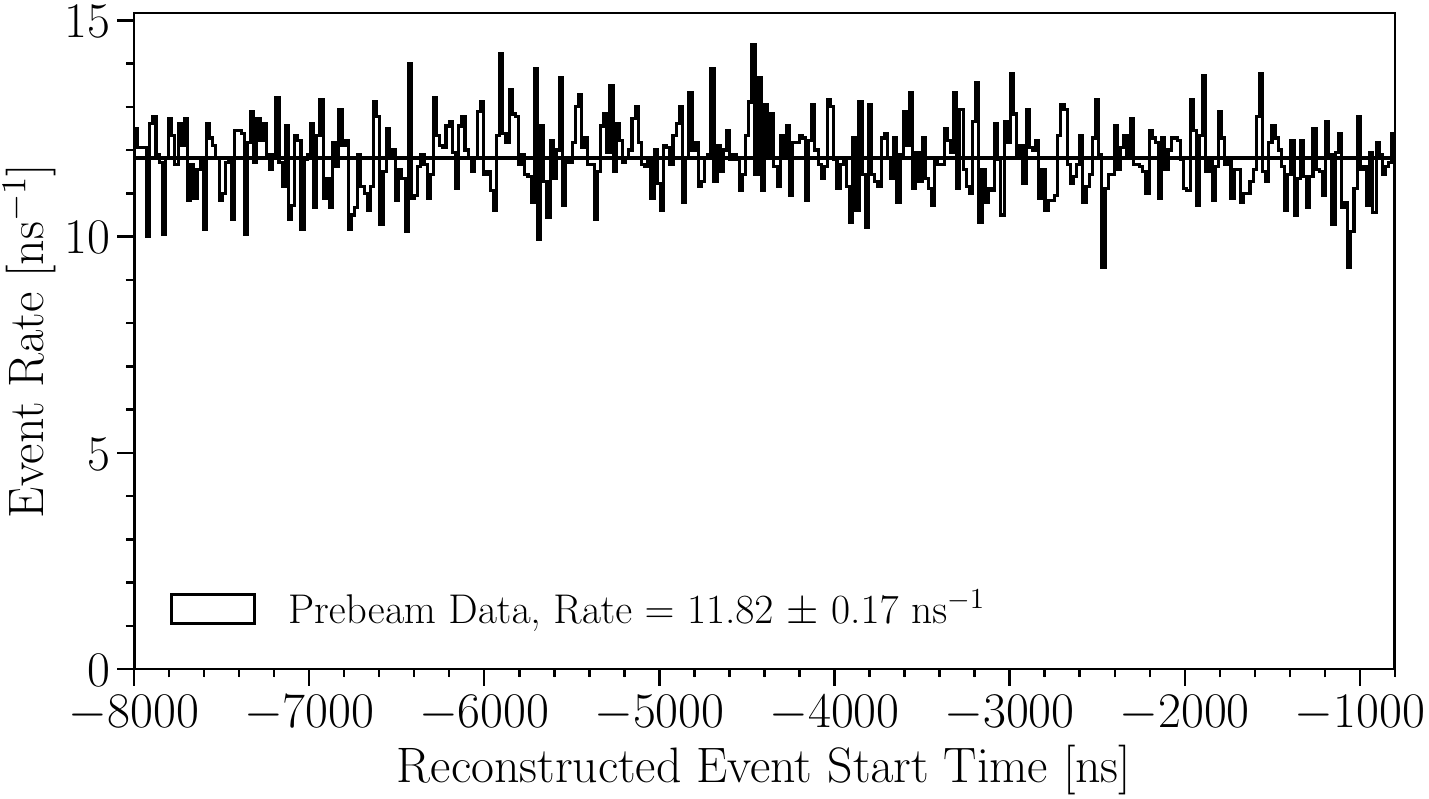}
  \caption{Time distribution of prebeam events after all selection criteria are applied. A fit to a constant rate yields an expected background of $11.82 \pm 0.17$ events per ns.}
  \label{fig:prebeam_time_after_cuts}
\end{figure}

Fig.~\ref{fig:prebeam_time_after_cuts} shows the time distribution of the surviving prebeam events after all selections have been applied. The prebeam time window is defined as reconstructed start times between $-8000~\mathrm{ns}$ and $-800~\mathrm{ns}$ relative to an external beam current monitor. The event rate, integrated over the full exposure of  $1.23 \times 10^{21}$ POT, is consistent with a time-independent background.

A fit with a constant function yields a background rate of $11.82 \pm 0.17$ events/ns. The fit quality is evaluated using Pearson's $\chi^2$ test, giving $\chi^2 = 427.76$ for 399 degrees of freedom, corresponding to a reduced $\chi^2$ of 1.07. The result is consistent with a uniform background rate and shows no statistically significant residual time dependence following the event selection.

\begin{figure}[h]
  \centering
  \includegraphics[width=\linewidth]{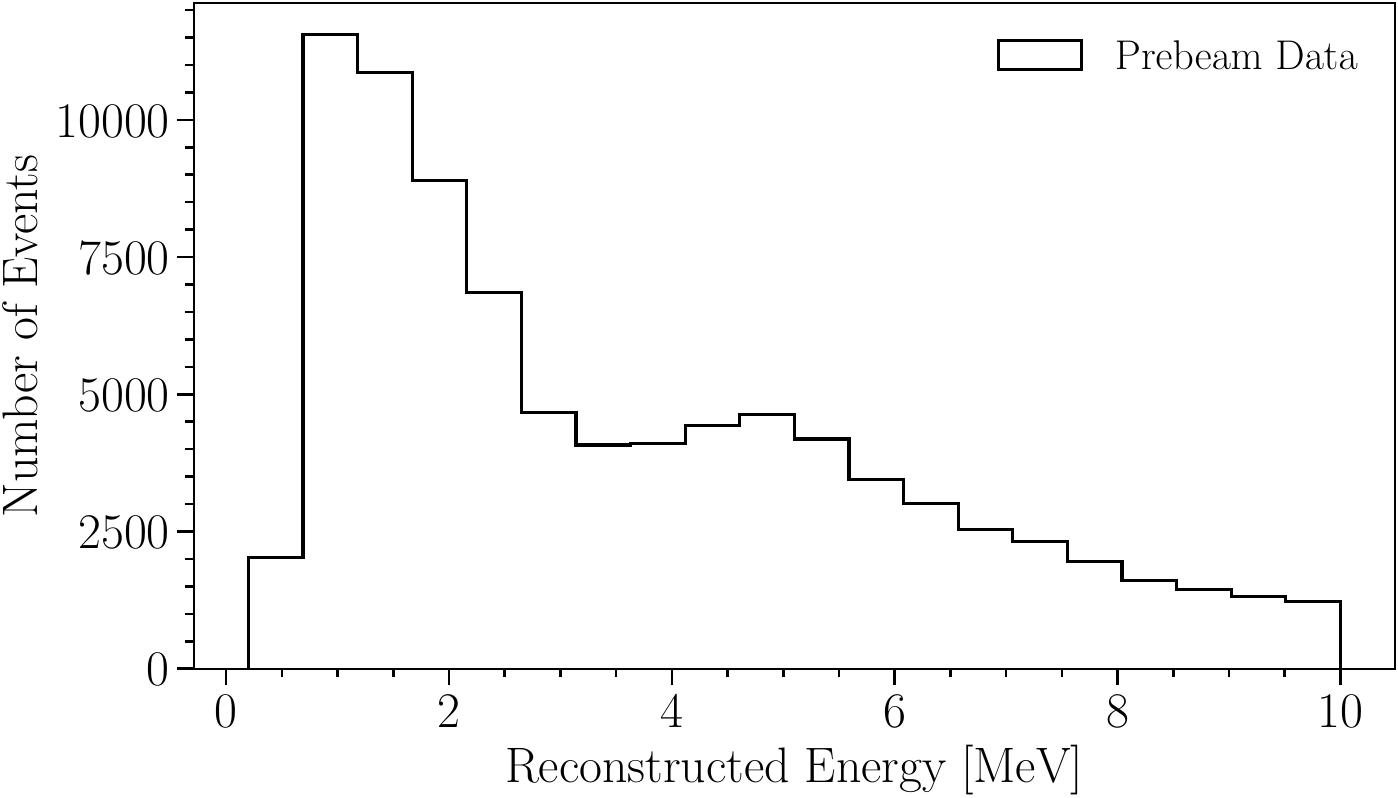}
  \caption{Reconstructed energy spectrum of prebeam events surviving the full event selection. The feature near 5~MeV may be associated with gamma rays from neutron capture on argon, although additional study is required to establish its origin.}
  \label{fig:prebeam_energy_after_cuts}
\end{figure}

The reconstructed energy spectrum of the selected prebeam sample is presented in Fig.~\ref{fig:prebeam_energy_after_cuts}. Most events cluster near 1~MeV, consistent with low-energy electromagnetic backgrounds that satisfy the Cherenkov-enhanced event selection.

A second feature is visible near 5~MeV. One possible explanation is gamma-ray emission following neutron capture on argon through the reaction $^{40}\text{Ar}(n,\gamma)^{41}\text{Ar}$, which produces de-excitation gamma rays with energies totaling up to approximately 6.1~MeV. These neutrons could originate from previous beam spills and remain in the experimental hall before thermalizing and capturing in the detector. Although this hypothesis is compatible with the observed energy scale, dedicated studies are needed to determine the origin of this feature.

\begin{figure}[h]
  \centering
  \includegraphics[width=\linewidth]{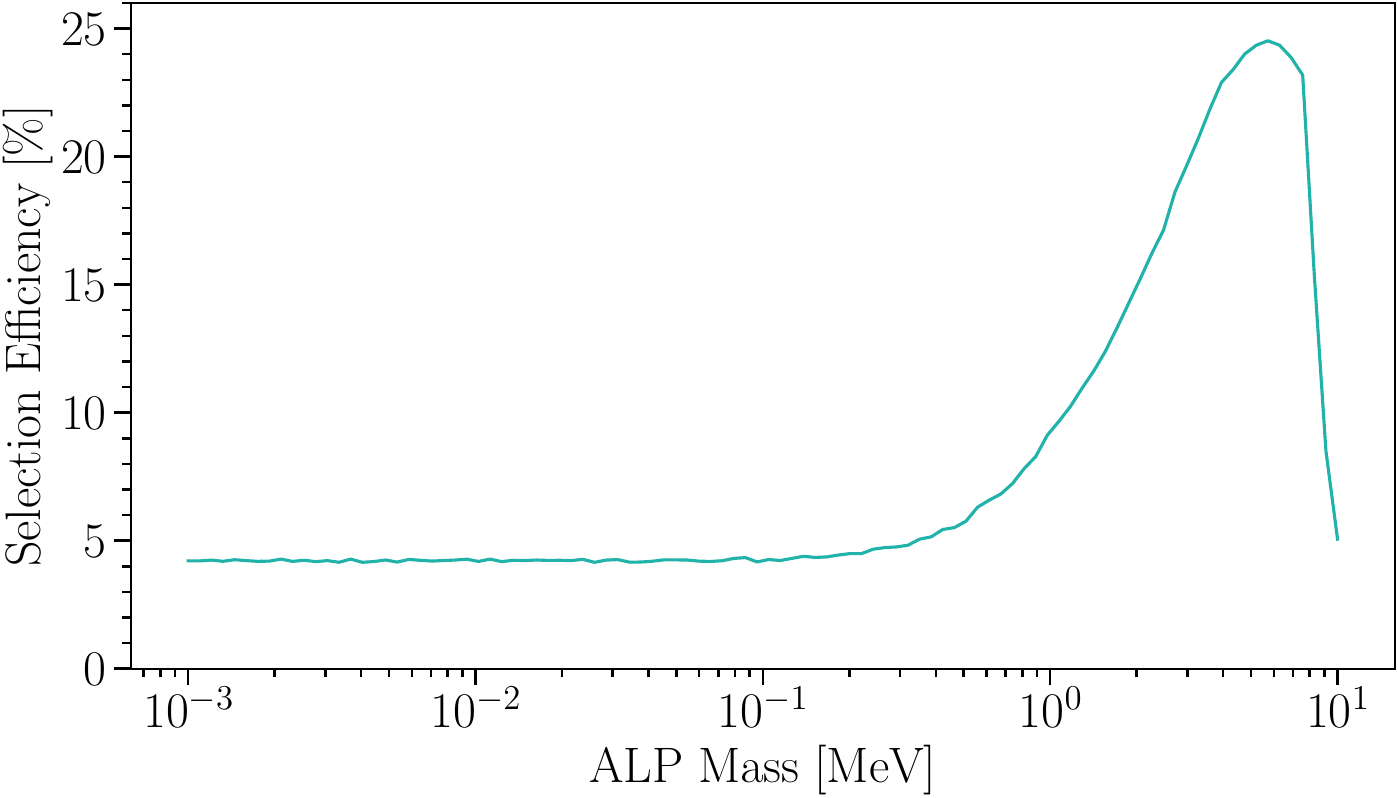}
  \caption{Overall signal selection efficiency as a function of ALP mass. The efficiency increases with mass, reaching approximately 25\% near 6~MeV before falling rapidly because of the upper reconstructed energy requirement of 10~MeV.}
  \label{fig:alp_mass_eff}
\end{figure}

The overall signal efficiency after all selection requirements depends strongly on the ALP mass and is shown in Fig.~\ref{fig:alp_mass_eff}. For ALP masses below approximately 0.3~MeV, the efficiency remains nearly constant at about 4\%, reflecting the small visible energy deposited by these events. As the ALP mass increases, the larger electromagnetic energy deposition improves reconstruction and event selection, causing the efficiency to rise steadily. The efficiency reaches a maximum of roughly 25\% near 6~MeV before decreasing sharply at higher masses, where an increasing fraction of signal events fail the upper reconstructed energy requirement of 10~MeV.

\section{\label{sec:results}Axion-Like Particle Fit}
The fit is performed in the two-dimensional space of reconstructed energy and event start time. Including timing information enhances the separation of beam-correlated ALP signals from the uniform steady-state background. A frequentist scan is performed over a grid of $10^4$ ALP parameter points, in addition to the background-only hypothesis. At each point, simulated ALP events are propagated through the full detector simulation, reconstruction, and event selection to produce signal templates. These templates are compared to the observed data by minimizing an effective binned likelihood that accounts for finite Monte Carlo statistics~\cite{Arguelles:2019izp}. Confidence intervals are derived using Wilks' theorem~\cite{Wilks:1938dza}.
 
\subsection{Fitting Procedure}
The likelihood fit is performed using two-dimensional templates in reconstructed energy and event start time. The energy axis is divided into 12 uniform bins spanning 0.2 to 10~MeV, while the time axis consists of 20 uniform bins covering $-600$~ns to $-424$~ns relative to an external beam current monitor. This time period defines the physics region of interest (ROI), beginning with the earliest possible arrival of relativistic particles at the detector and ending with the onset of neutron backgrounds. The start of the ROI is determined using an external EJ-301 scintillator detector to monitor beam-related activity in the experimental hall. By identifying the statistically significant increase in activity, the arrival time of prompt relativistic particles is established. Pulse-level activity distributions in the CCM200 detector are then used to identify the onset of neutron backgrounds, which defines the end of the ROI. For further details, see Ref.~\cite{CCM:2021leg}.  

\begin{figure}[h]
  \centering
  \includegraphics[width=\linewidth]{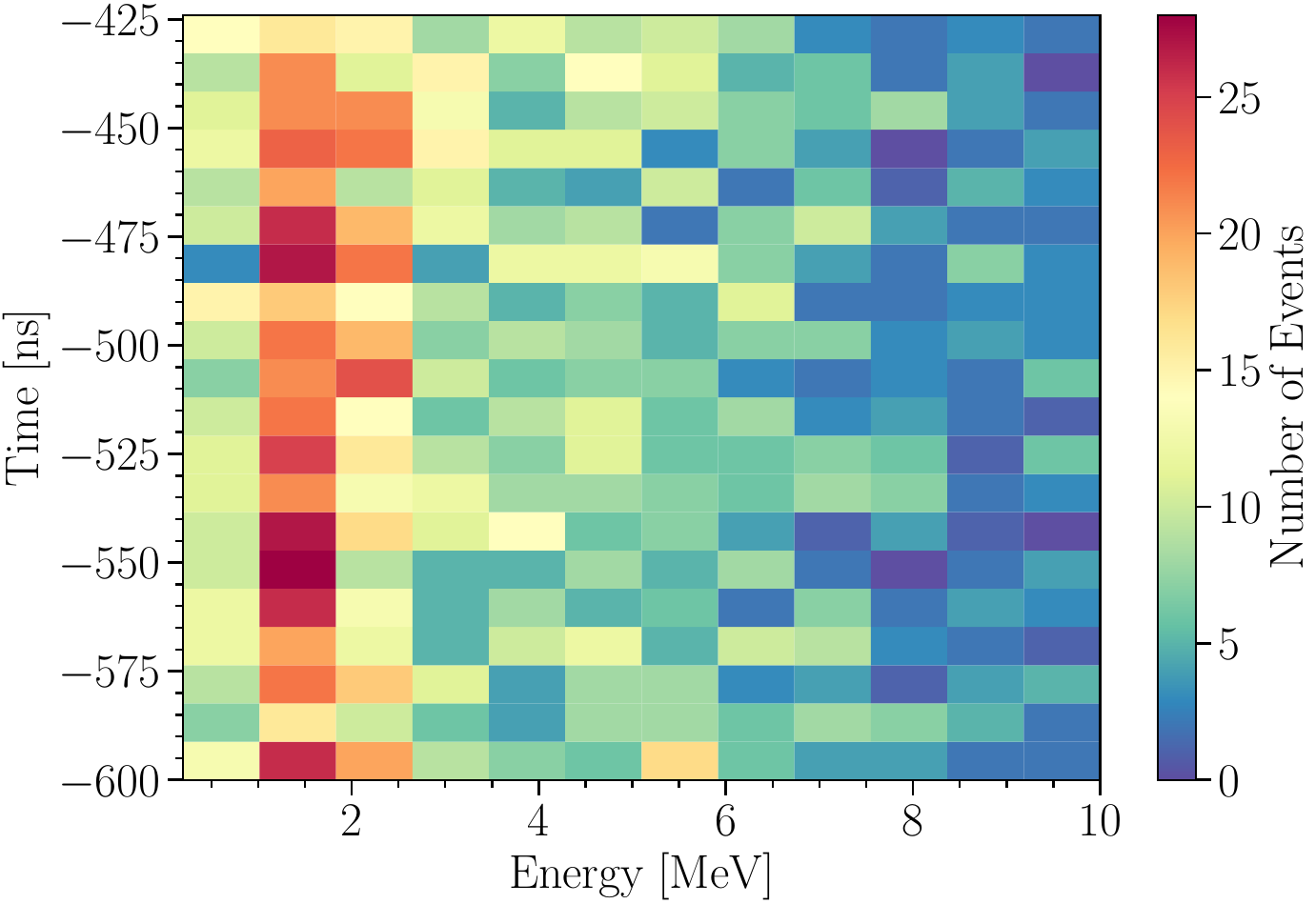}
  \caption{Two-dimensional reconstructed energy and event start-time distribution of the observed data.}
  \label{fig:fitting_energy_time}
\end{figure}

Observed data events in the time ROI and the energy range of the analysis are filled directly into the two-dimensional $(E,t)$ histogram, demonstrated in Fig.~\ref{fig:fitting_energy_time}. To reduce statistical fluctuations in the signal and background templates, the energy distributions are obtained from simulated ALP or prebeam events, while the time distributions are modeled analytically and normalized to the corresponding energy spectra. This hybrid approach minimizes the impact of finite Monte Carlo statistics on the time dimension.

The selected prebeam sample is consistent with a constant event rate in time. Consequently, the expected background time distribution is modeled with a uniform PDF.

The expected signal time distribution is constructed by convolving the simulated event time offsets, which account for the ALP time-of-flight to the detector along with reconstruction-level timing offsets, with the time profile of the beam pulse. The beam pulse has a characteristic triangular time profile that is 290~ns wide at the base, shown in Ref.~\cite{Newmark:2026sea}. For each simulated event, the probability of falling within a given time bin is computed by integrating the beam time profile over the bin boundaries, and these probabilities are summed using the corresponding event weights. 

\subsection{Sources of Systematic Uncertainties}
As a proof-of-concept analysis, the available statistics are limited, and the measurement is therefore dominated by statistical uncertainties. Nevertheless, several sources of systematic uncertainty are considered and incorporated into the likelihood through nuisance parameters that are profiled during the fit. The dominant contributions arise from the steady-state background normalization, the total accumulated POT, and the absolute beam timing.

\subsubsection{Background Normalization}
The selected prebeam sample exhibits a time-independent event rate (Fig.~\ref{fig:prebeam_time_after_cuts}), with a best-fit value of $11.82 \pm 0.17$ events/ns. The background normalization is therefore treated as a nuisance parameter with a Gaussian prior centered at 11.82 events/ns and a standard deviation of 0.17 events/ns. This allows the fit to vary the overall background level within the measured uncertainty while constraining deviations from the nominal expectation.

\subsubsection{Total Protons on Target}
The expected ALP signal scales linearly with the total POT accumulated during data taking. The POT is determined from a calibration between an external beam current monitor and the beam current reported by the LANSCE accelerator division, resulting in an overall normalization uncertainty of approximately 5\%. This uncertainty is incorporated through a multiplicative signal normalization parameter with a Gaussian prior centered at unity and a standard deviation of 0.05.

\subsubsection{Beam Timing}
The time distribution of the signal prediction depends on the relative timing between the proton beam and the detector. The earliest arrival time of prompt beam related particles is known to within approximately 30~ns, due to statistics of the ROI start time determination, producing a corresponding uncertainty in the predicted ALP timing distribution. To account for this effect, the beam start time is allowed to vary during the fit with a Gaussian prior centered at -600~ns and a width of 30~ns.

\subsection{Fit Results}
The fit is performed by minimizing the negative log-likelihood while profiling the three nuisance parameters introduced in the previous section. The effective likelihood, including Gaussian pull terms for these nuisance parameters, is given in Eq.~(\ref{eq:leff}). This formalism depends on the observed counts $k$ and the weighted Monte Carlo quantities $\alpha = \frac{(\sum_i w_i)^2}{\sum_i w_i^2} + 1$ and $\beta = \frac{\sum_i w_i}{\sum_i w_i^2}$, where $w_i$ is the weight assigned to each simulated event~\cite{Arguelles:2019izp}. The pull terms correspond to Gaussian priors centered at $\mu_i$ with standard deviations $\sigma_i$ for each nuisance parameter $x_i$.

\begin{equation}
\begin{split}
\mathcal{L}_{\text{eff}}(k, \alpha, \beta)
    &= \prod_{(E,t)} \frac{\Gamma(k+\alpha)}
            {k!~\Gamma(\alpha)}
       \frac{\beta^{\alpha}}
            {(1+\beta)^{k+\alpha}} \\
    &\quad \prod_i
    \exp\left(
    -\frac{(x_i-\mu_i)^2}{2\sigma_i^2}
    \right)
\end{split}
\label{eq:leff}
\end{equation}

Under the background-only hypothesis, the fit returns a negative log-likelihood of 578.28 and a profiled background rate of 11.74 events/ns, consistent with the measured value within its $1\sigma$ uncertainty range.

Allowing for an ALP signal, the best-fit point occurs at $m_a = 0.18$~MeV and $g_{a\gamma} = 1.63 \times 10^{-4}~\text{GeV}^{-1}$, with a negative log-likelihood of 577.34. The profiled nuisance parameters remain close to their nominal values, yielding a POT scaling of 0.99, a beam start time of -584.52 ns, and a background rate of 11.67 events/ns. The small shifts in these parameters indicate that the data are well described without requiring substantial deviations from the nominal model.

\begin{figure}[h]
    \centering
    \includegraphics[width=\linewidth]{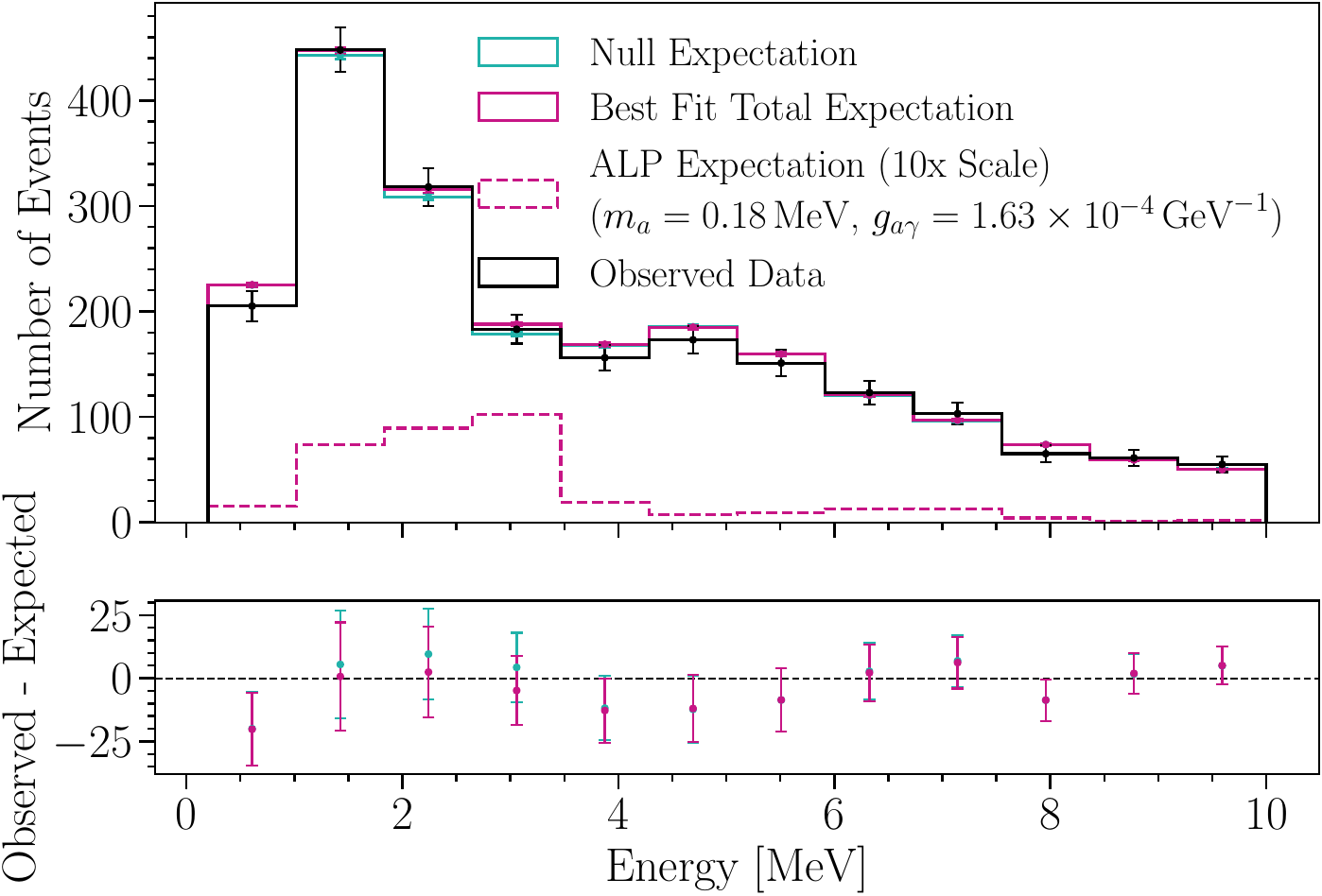}
    \caption{Projected reconstructed energy distribution for the observed data (black), the background-only fit (cyan), and the best-fit signal-plus-background model (magenta). The best fit ALP signal is at $m_a = 0.18$~MeV and $g_{a\gamma} = 1.63 \times 10^{-4}~\text{GeV}^{-1}$. The dashed magenta line shows the ALP expectation at 10x scale for visibility. The lower panel shows the residuals relative to the observed data.}
    \label{fig:fitting_energy}
\end{figure}

\begin{figure}[h]
    \centering
    \includegraphics[width=\linewidth]{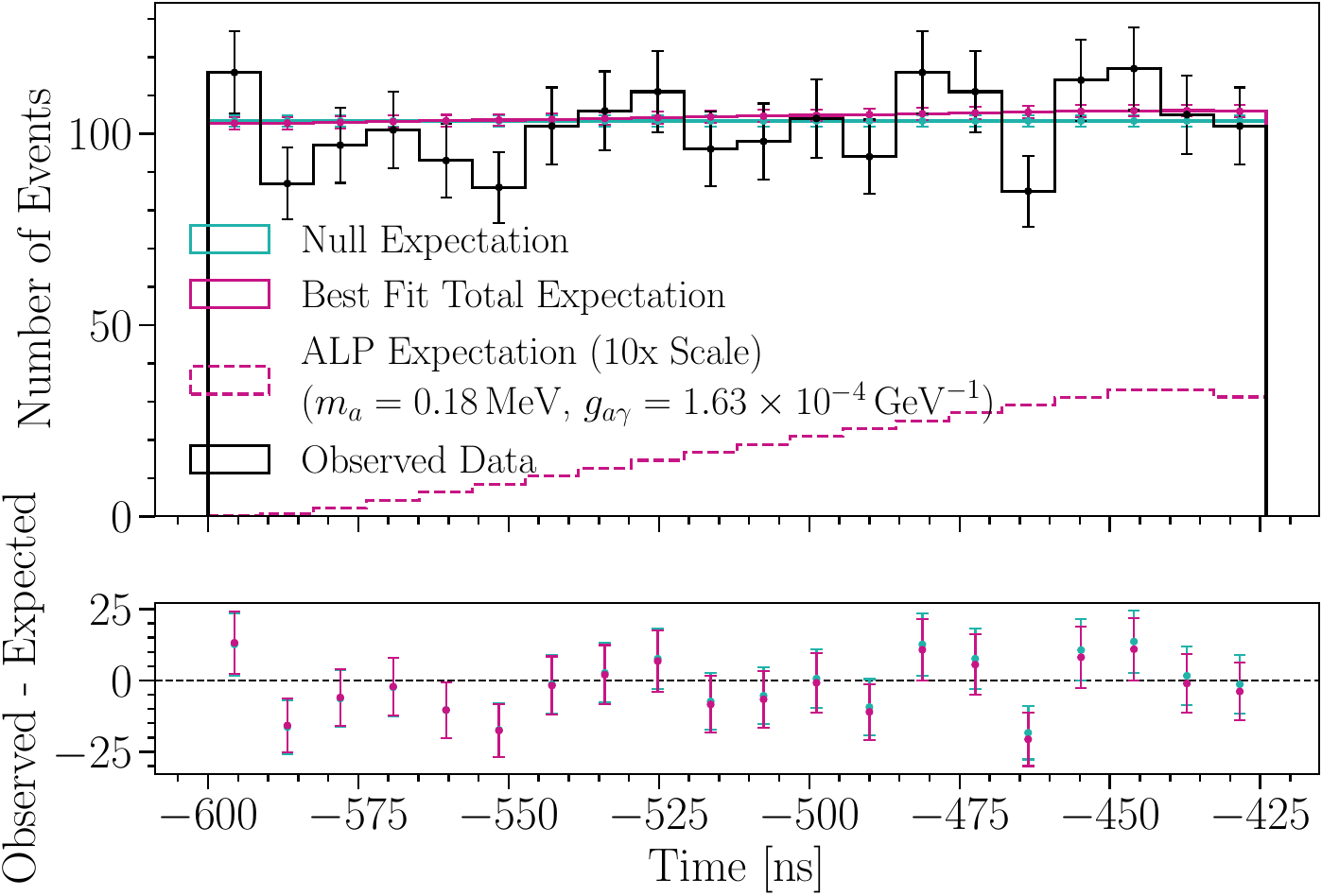}
    \caption{Projected event start-time distribution for the observed data (black), the background-only fit (cyan), and the best-fit signal-plus-background model (magenta). The best fit ALP signal is at $m_a = 0.18$~MeV and $g_{a\gamma} = 1.63 \times 10^{-4}~\text{GeV}^{-1}$. The dashed magenta line shows the ALP expectation at 10x scale for visibility. The lower panel shows the residuals relative to the observed data.}
    \label{fig:fitting_time}
\end{figure}

The best-fit signal yields a decrease in the negative log-likelihood of 0.94 relative to the background-only hypothesis. Under Wilks' theorem, for two degrees of freedom corresponding to the ALP mass and coupling, this corresponds to a local significance of $0.86\sigma$. The observed improvement is therefore compatible with a statistical fluctuation, and no evidence for ALP production is found.

\begin{table*}[t]
    \caption{Comparison between CCM120 and CCM200 (this work). The prebeam efficiency defines the rejection of steady state backgrounds, discussed in Sec.~\ref{sec:event_selection}. The efficiency for $m_a = 1~\mathrm{MeV}$ ALPs between the two analyses is provided for comparison purposes at one point in the parameter space.}
    \begin{ruledtabular}
        \begin{tabular}{cccccc}
        Detector Configuration & Hybrid Detection & Photocathode Coverage & Exposure & Prebeam Eff. & 1~MeV ALP Eff. \\
        \hline
        CCM120 & No & 30\% & $1.79\times10^{21}$ POT & 3.2\% & 19.0\% \\
        CCM200 & Yes & 50\% & $1.23\times10^{21}$ POT & 0.49\%  & 9.22 \% \\
        \end{tabular}
    \end{ruledtabular}
    \label{table:ccm120_vs_ccm200}
\end{table*}

Fig.~\ref{fig:fitting_energy} shows the reconstructed energy projection of the fit along with the observed data in black. The background-only prediction is shown in cyan, while the best-fit signal-plus-background model is shown in magenta, with the dashed curve indicating the signal component alone. Note that the signal-only component (dashed magenta) is magnified by a factor of 10 to show the features more clearly. The preferred signal produces a modest excess between approximately 1 and 3 MeV, but the statistical uncertainty in this region is sufficiently large that the improvement in fit quality is minimal.

The corresponding event start-time projection is shown in Fig.~\ref{fig:fitting_time}. The background expectation is uniform across the analysis window, whereas the signal model exhibits the characteristic triangular timing profile produced by the proton beam pulse, beginning near -584 ns and extending over approximately 290 ns. The observed timing distribution shows no statistically significant preference for this signal-like structure.

\begin{figure}[h]
  \centering
  \includegraphics[width=\linewidth]{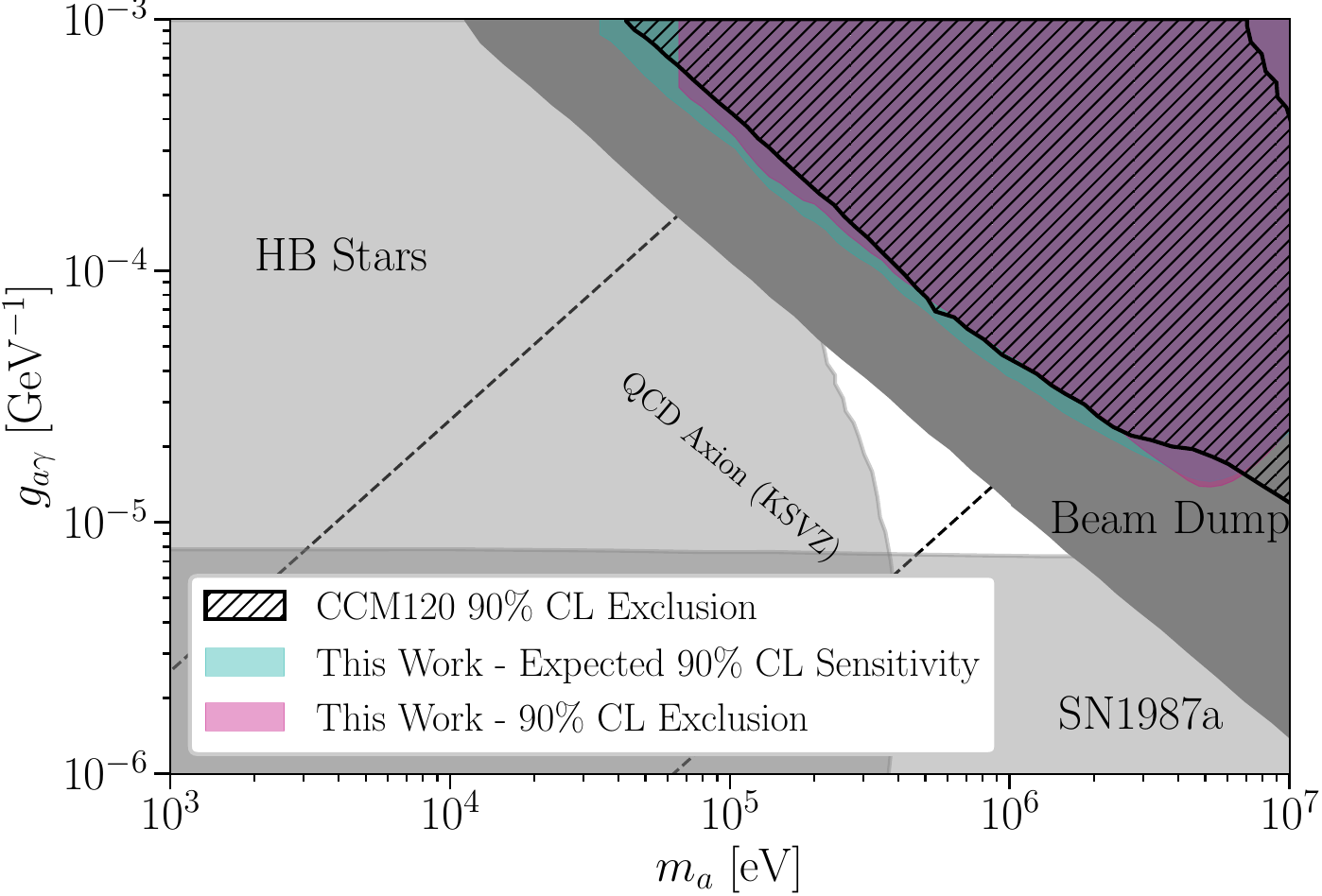}
  \caption{Observed (magenta) and expected (cyan) 90\% confidence level exclusion limits on the ALP parameter space obtained in this analysis. No statistically significant evidence for an ALP signal is observed. Despite using only approximately 70\% of the proton exposure of the previous CCM120 search, this analysis extends the excluded parameter space through improved background rejection enabled by Cherenkov-based event discrimination.}
  \label{fig:actual_sensitivity}
\end{figure}

Finally, Fig.~\ref{fig:actual_sensitivity} presents the observed 90\% confidence level exclusion limits together with the expected sensitivity, where the 90\% confidence level corresponds to $\Delta\mathrm{NLLH} = 2.30$. In addition to the previous CCM120 search for ALPs, the allowed parameter space for the KSVZ QCD axion model is denoted by the dashed black lines~\cite{KSVZ_1,KSVZ_2}. Existing constraints on the ALP parameter space are also shown for comparison. Beam dump experiments provide the strongest laboratory limits in this region by probing similar ALP production and decay processes at comparable masses and couplings~\cite{JAECKEL2016482,doi:10.1142/S0217751X9200171X}. Astrophysical constraints are also considered, including bounds from energy loss arguments in horizontal branch (HB) stars and observations of the neutrino signal from SN1987a~\cite{PhysRevD.75.013004,PhysRevLett.93.171104,PhysRevLett.97.151802,PhysRevLett.98.131802,PhysRevLett.98.050402,PhysRevLett.99.121103,DeRocco:2020xdt,Masso_2005}. These limits provide strong complementary coverage, especially at low masses and couplings, but depend on assumptions about stellar models and ALP interactions in astrophysical environments.

Although this analysis uses approximately 70\% of the integrated POT compared to the previous CCM120 search, it excludes additional regions of ALP parameter space due to increased suppression of steady state backgrounds. This demonstrates the improved sensitivity achieved through the hybrid Cherenkov-scintillation detection and its enhanced background rejection.

\section{\label{sec:conclusion}Summary and Future Prospects}
This analysis presents the first search for axion-like particles using a 10-ton liquid argon hybrid Cherenkov-scintillation optical detector in a beam dump environment. By exploiting four complementary observables based on the distinct timing, wavelength response, angular structure, and topology of Cherenkov and scintillation light, this work demonstrates the ability of hybrid optical detection to enhance discrimination between electromagnetic signal events and beam-related backgrounds.

Table~\ref{table:ccm120_vs_ccm200} compares the detector configurations and analysis performance of the CCM120 ALP search~\cite{CCM:2021jmk} and this work. Despite using a smaller exposure, this work achieves a substantially lower prebeam background efficiency, with a factor of $\sim$6 improvement compared to CCM120. This improvement is enabled by the hybrid Cherenkov-scintillation detection capabilities of CCM200, which provide additional event discrimination beyond scintillation information alone. The improved background rejection compensates for the reduced exposure and signal efficiency, resulting in comparable or improved ALP sensitivity across the presented mass range.

The sensitivity of the current search is optimized for low-energy signals, reflecting the detector calibration based on the $^{22}$Na source presented in Ref.~\cite{CCM:2025dbq}. Ongoing calibration efforts using Michel electrons will extend the validated energy range to approximately 50~MeV, enabling future searches to probe higher-mass ALPs with improved sensitivity. The collaboration is also working on direction reconstruction that will be especially useful to isolate boosted signals expected from higher energy ALPs. Additionally, future analyses will utilize the full CCM200 physics dataset, comprising approximately $3 \times 10^{21}$ POT. 

Beyond this specific search, this work demonstrates that hybrid optical detectors can substantially improve sensitivity for future accelerator-based searches for weakly interacting particles. 

\begin{acknowledgments}
We acknowledge the support of the Los Alamos National Laboratory LDRD and the U.S. Department of Energy Office of Science funding. We also wish to acknowledge support from the LANSCE Lujan Center and LANL’s Accelerator Operations and Technology (AOT) division. This research used resources provided by the Los Alamos National Laboratory Institutional Computing Program, which is supported by the U.S. Department of Energy National Nuclear Security Administration under Contract No.~89233218CNA000001. DAN is supported by the NSF Graduate Research Fellowship under Grant No.~2141064 and MIT School of Science. AAA-A, CFMA, JCD, and MCE acknowledge support from DGAPA-UNAM Grant No.~PAPIIT-IN104026. We also wish to thank the SubMIT supercomputer and physics analysis facility at MIT~\cite{acosta2026submit} for the generous use of their computing resources.
\end{acknowledgments}

\appendix
\section{\label{sec:app}Position and Energy Reconstruction}
Position reconstruction is performed using the \texttt{GraphNeT} machine learning framework~\cite{Sogaard:2022qgg}, which was adapted for the CCM200 detector geometry and achieves a spatial resolution of approximately 5~cm in each dimension. The reconstructed position is then used together with the measured charge deposition to estimate the event energy, yielding an energy resolution of approximately 12\% at 1~MeV.

\subsection{Position Reconstruction}
Position reconstruction provides strong discrimination between signal and background events. Beam-related and cosmic-ray backgrounds predominantly originate outside the detector and therefore reconstruct near its boundaries, while signal events are expected to be uniformly distributed throughout the active volume. Improving the vertex resolution directly enhances the signal-to-background ratio.

To improve upon the charge-weighted PMT position based reconstruction used in previous CCM120 analyses, a graph neural network based position reconstruction algorithm was developed using the open-source \texttt{GraphNeT} framework~\cite{Sogaard:2022qgg}. The model employs a \texttt{DeepIce} transformer backbone~\cite{deepice}, which naturally handles the irregular geometry and sparse optical readout of the detector.

The network is trained on reconstructed photoelectron pulses, using the pulse time, charge, PMT position, and PMT coating status as input features. Embedded pulse features are processed by transformer layers with self-attention, allowing the network to learn the spatial and temporal correlations of detected light. Event-level representations are formed using a learned classification token, whose output is passed to a regression head that predicts the interaction vertex.

Training is performed with the LogCosh loss function, detailed in Eq.~(\ref{eq:logcosh}). In this loss function, $\Delta x_i = \hat{y}_i - y_i$, for $y_i$ representing the true position and $\hat{y}_i$ is the model prediction. This loss was optimized using the \texttt{Adam} optimizer~\cite{adamoptimizer}.

\begin{equation}
    \text{LogCosh}(\Delta x_i) = \sum_{i=1}^{n} \left[ \Delta x_i + \ln(1 + e^{-2\Delta x_i}) - \ln(2) \right]
\label{eq:logcosh}
\end{equation}

The position reconstruction model is trained on approximately $6\times10^6$ simulated electron events generated uniformly throughout the detector volume, including the veto region, with isotropic directions and energies uniformly distributed between 0.1 and 30~MeV. The training target is the true electron injection position. To accurately reproduce detector response, the simulation incorporates the full data acquisition chain, including measured board-to-board timing offsets, PMT electron transit time calibrations, and 2~ns waveform digitization. A data-driven noise model is then overlaid by sampling prebeam pulse series from physics data, reproducing the steady-state PMT noise observed during data taking.

The simulated events are processed using the same event reconstruction chain as the physics data to avoid biasing the selection. Additionally, the reconstructed pulse series are rebinned to 6~ns time bins to reduce data quantity. To improve robustness against detector systematics, each training event includes random variations corresponding to the measured 10\% uncertainty in the scintillation light yield~\cite{CCM:2025dbq} and a small ($\sigma=0.5$~ns) Gaussian time jitter applied to individual pulse times, reducing sensitivity to digitization artifacts.

\subsubsection{Position Reconstruction Performance}
\begin{figure}[h]
  \centering
  \includegraphics[width=\linewidth]{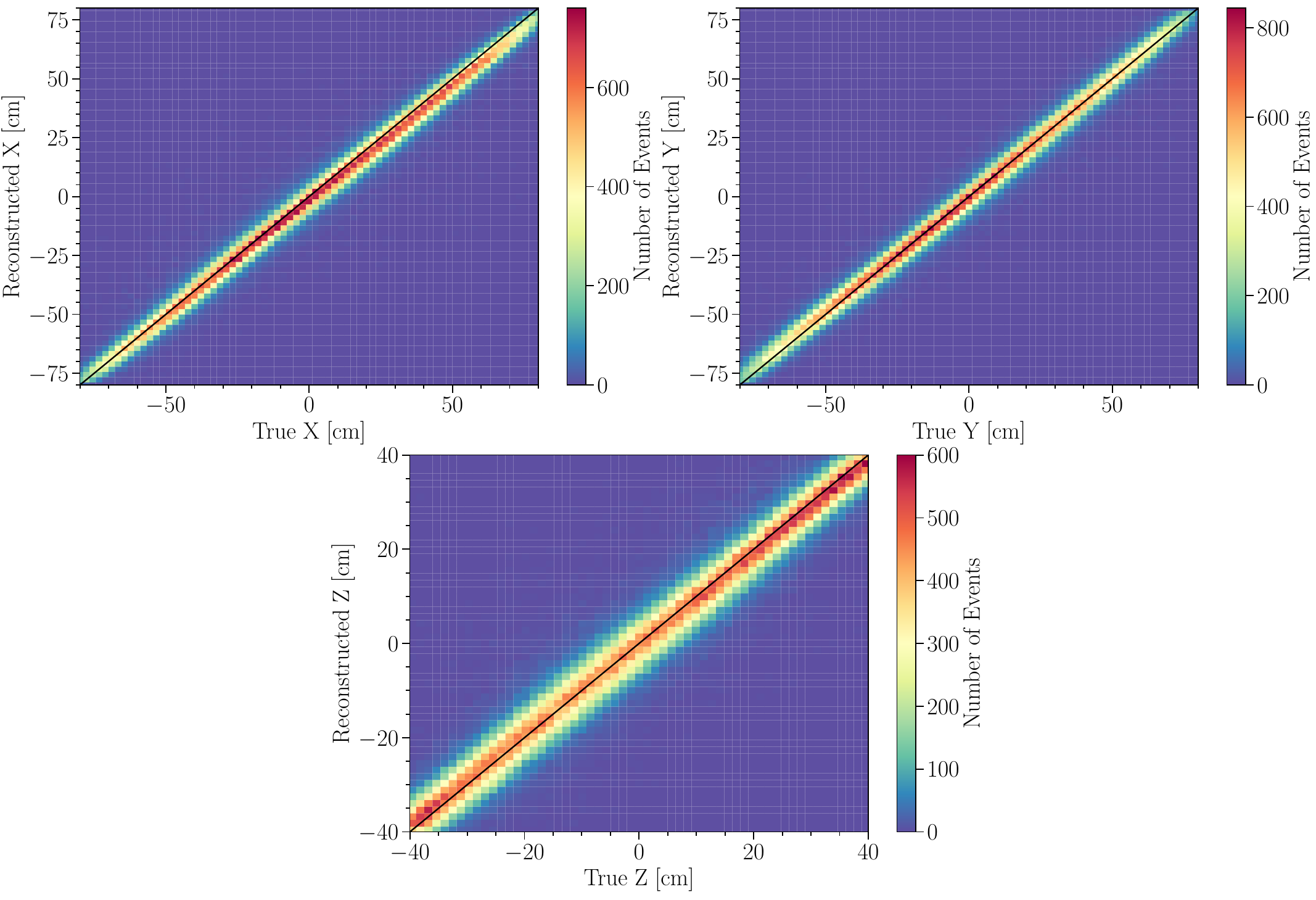}
  \caption{Two-dimensional distributions comparing the true and reconstructed event positions for the \texttt{GraphNeT} position reconstruction model. The $x$, $y$, and $z$ coordinates are shown in the top left, top right, and bottom panels, respectively. The diagonal line denotes the ideal reconstruction.}
  \label{fig:true_vs_reco_xyz}
\end{figure}

The model is evaluated on a held-out validation sample comprising 10\% of the simulated events. Fig.~\ref{fig:true_vs_reco_xyz} compares the reconstructed and true interaction positions in each spatial dimension after applying the standard fiducial selection ($r<80$~cm and $|z|<40$~cm). The reconstructed positions exhibit a strong linear correlation with the true positions, with no significant bias observed across the fiducial volume.

\begin{figure}[h]
  \centering
  \includegraphics[width=\linewidth]{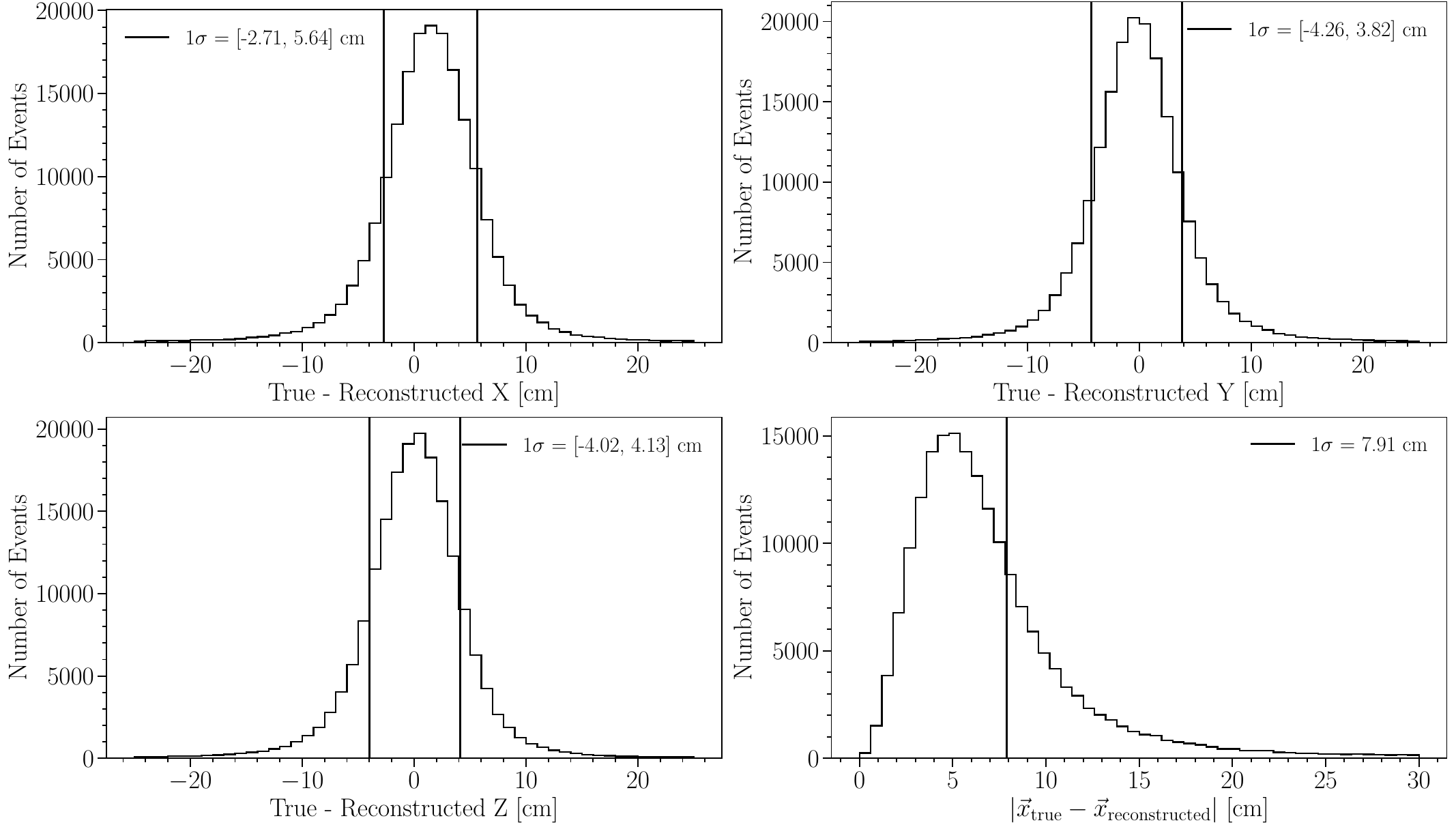}
  \caption{Residual distributions for the reconstructed event position. The $x$, $y$, and $z$ coordinates each exhibit a resolution of approximately 5~cm at the $1\sigma$ level, while the combined three-dimensional position resolution is 7.91~cm.}
  \label{fig:resolution_xyzalldir}
\end{figure}

The corresponding residual distributions, shown in Fig.~\ref{fig:resolution_xyzalldir}, are used to quantify the spatial resolution. The reconstruction achieves an approximately isotropic resolution of $\sim$5~cm in each coordinate, corresponding to a three-dimensional position resolution of 7.9~cm at the $1\sigma$ level. This represents roughly a four-fold improvement over the charge-weighted reconstruction previously used in CCM120 analyses, enabling more precise fiducialization and improved rejection of backgrounds originating near the detector boundaries.

\begin{figure}[h]
  \centering
  \includegraphics[width=\linewidth]{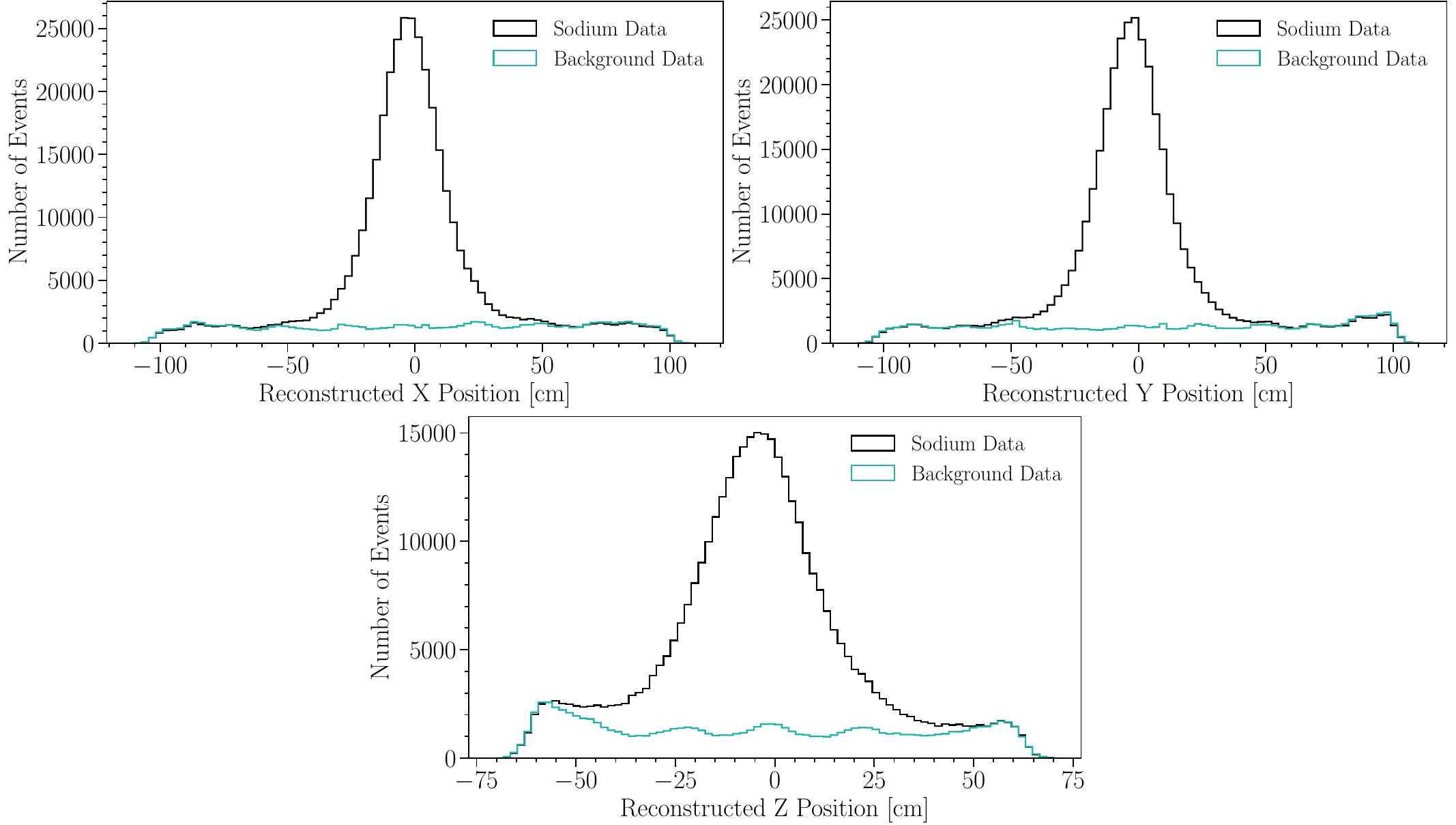}
  \caption{Reconstructed event positions for the $^{22}$Na calibration and background datasets. The calibration sample (black) is concentrated near the detector origin, as expected, whereas the background events (cyan) are distributed approximately uniformly within the active volume.}
  \label{fig:sodium_data}
\end{figure}

The position reconstruction is validated using $^{22}$Na calibration data collected with the source deployed at the detector center. Fig.~\ref{fig:sodium_data} compares the reconstructed position distributions for source data and a background-only dataset acquired with the source removed. As expected, the source data exhibit a pronounced peak near the detector origin, while the background sample is approximately uniform throughout the detector volume.

\begin{figure}[h]
  \centering
  \includegraphics[width=\linewidth]{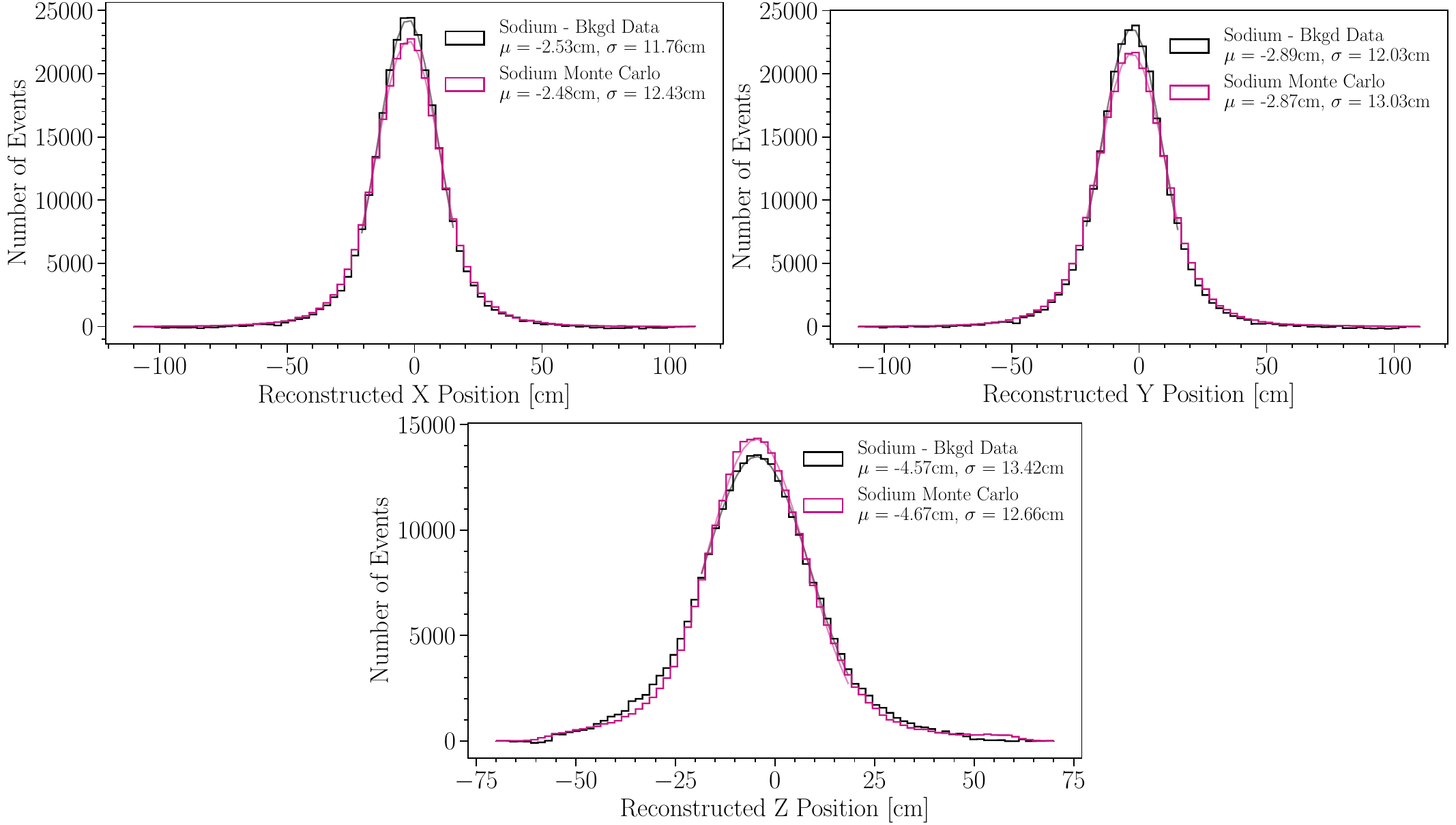}
  \caption{Reconstructed position distributions for background-subtracted $^{22}$Na calibration data and Monte Carlo simulation. Gaussian fits are used to extract the reconstructed source position and width, with agreement between data and simulation at the $\mathcal{O}(1~\mathrm{cm})$ level in all dimensions.}
  \label{fig:sodium_datavsmc}
\end{figure}

To isolate the $^{22}$Na contribution, the background-only sample is normalized to the same exposure and subtracted from the source data. The resulting distributions are compared with Monte Carlo simulation in Fig.~\ref{fig:sodium_datavsmc}. Gaussian fits to the reconstructed positions show agreement between data and simulation at the $\mathcal{O}(1~\mathrm{cm})$ level in all three spatial dimensions. The observed widths, of order 10~cm, are larger than the expected position resolution obtained from simulated electron events because the $^{22}$Na source emits gamma rays, which travel finite distances before producing secondary electrons. The agreement between data and simulation demonstrates that the reconstruction accurately reproduces the source position and detector response in calibration data.

\subsection{Energy Reconstruction}
In addition to position, energy reconstruction is also essential for event characterization. The deposited energy is inferred from the observed scintillation light, with the total number of detected photoelectrons (PEs) serving as a proxy for the true energy deposition. However, the light yield depends on the interaction position within the detector because of variations in light collection efficiency, optical propagation, and PMT response.

To account for these effects, the energy reconstruction incorporates the reconstructed interaction position together with the observed charge to estimate the deposited energy. This position-dependent approach significantly improves upon the uniform charge-to-energy calibration used in previous CCM120 analyses.

The deposited energy is estimated from the charge collected within the first 90~ns of an event using a position-dependent calibration. A large sample of simulated electrons, uniformly distributed throughout the detector with energies between 0.1 and 30~MeV, is used to derive this calibration. After reconstructing each interaction vertex using the techniques described previously, the detector is divided into 8~cm voxels, and the median charge-to-energy ratio ($Q/E$) is computed in each voxel. A spline interpolation is then used to obtain a continuous energy calibration as a function of reconstructed position.

\begin{figure}[h]
\centering
    \includegraphics[width=\linewidth]{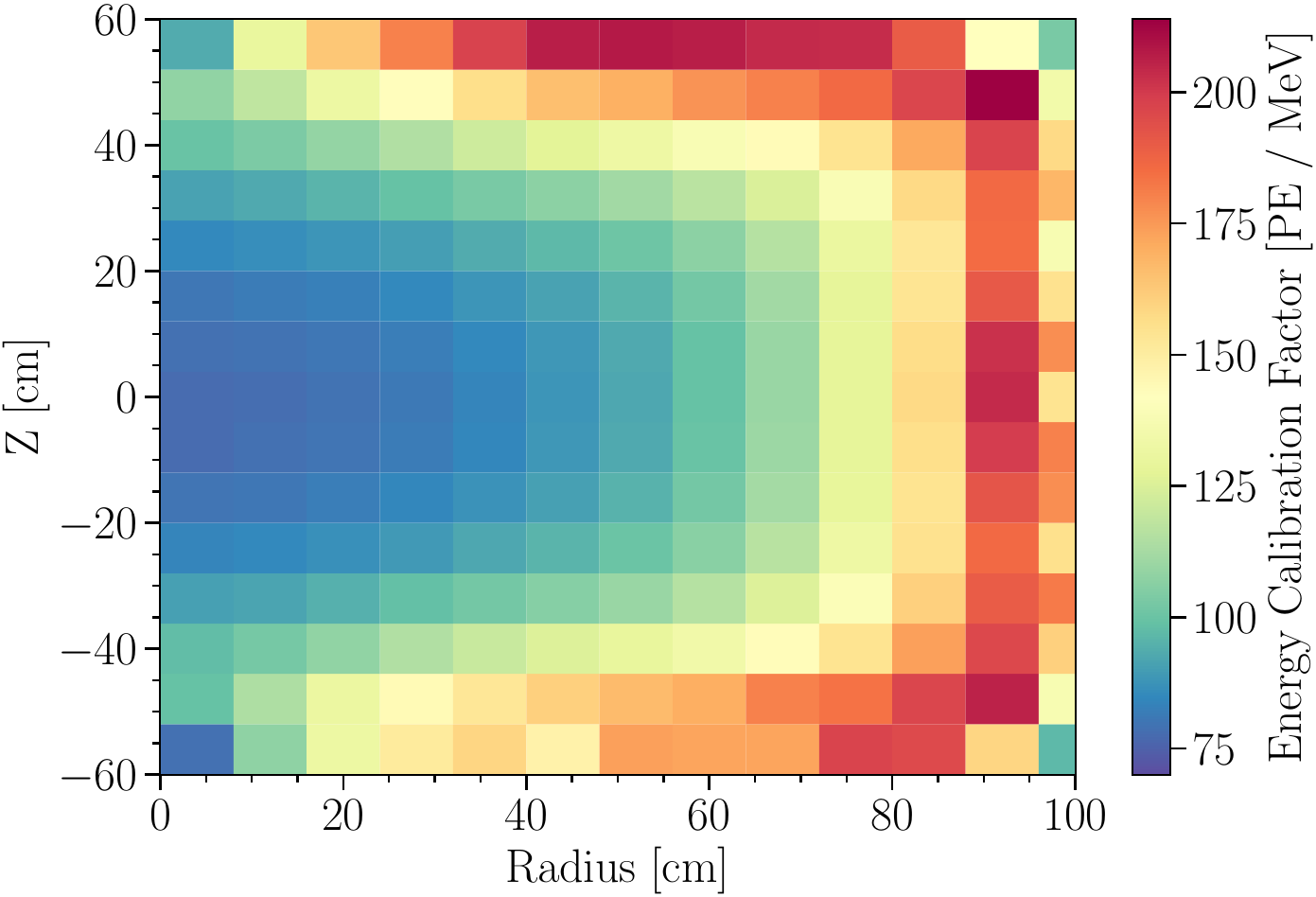}
    \caption{Energy calibration factor as a function of reconstructed radius and $Z$ position. The detector response varies from approximately 80~PE/MeV at the center to 200~PE/MeV near the edges, motivating the use of a position-dependent energy reconstruction.}
\label{fig:spline_scaling_vs_position}
\end{figure}

The resulting calibration, shown in Fig.~\ref{fig:spline_scaling_vs_position}, varies smoothly across the detector, increasing from approximately 80~PE/MeV at the center to 200~PE/MeV near the detector boundaries because of improved light collection. This variation of more than a factor of two highlights the importance of accounting for position when reconstructing the deposited energy. Restricting the charge integration to the first 90~ns also reduces contamination from delayed light and pile-up, providing a robust energy estimate for the low-energy, contained events relevant to this analysis.
\subsubsection{Energy Reconstruction Performance}

\begin{figure}[h]
    \centering
    \includegraphics[width=\linewidth]{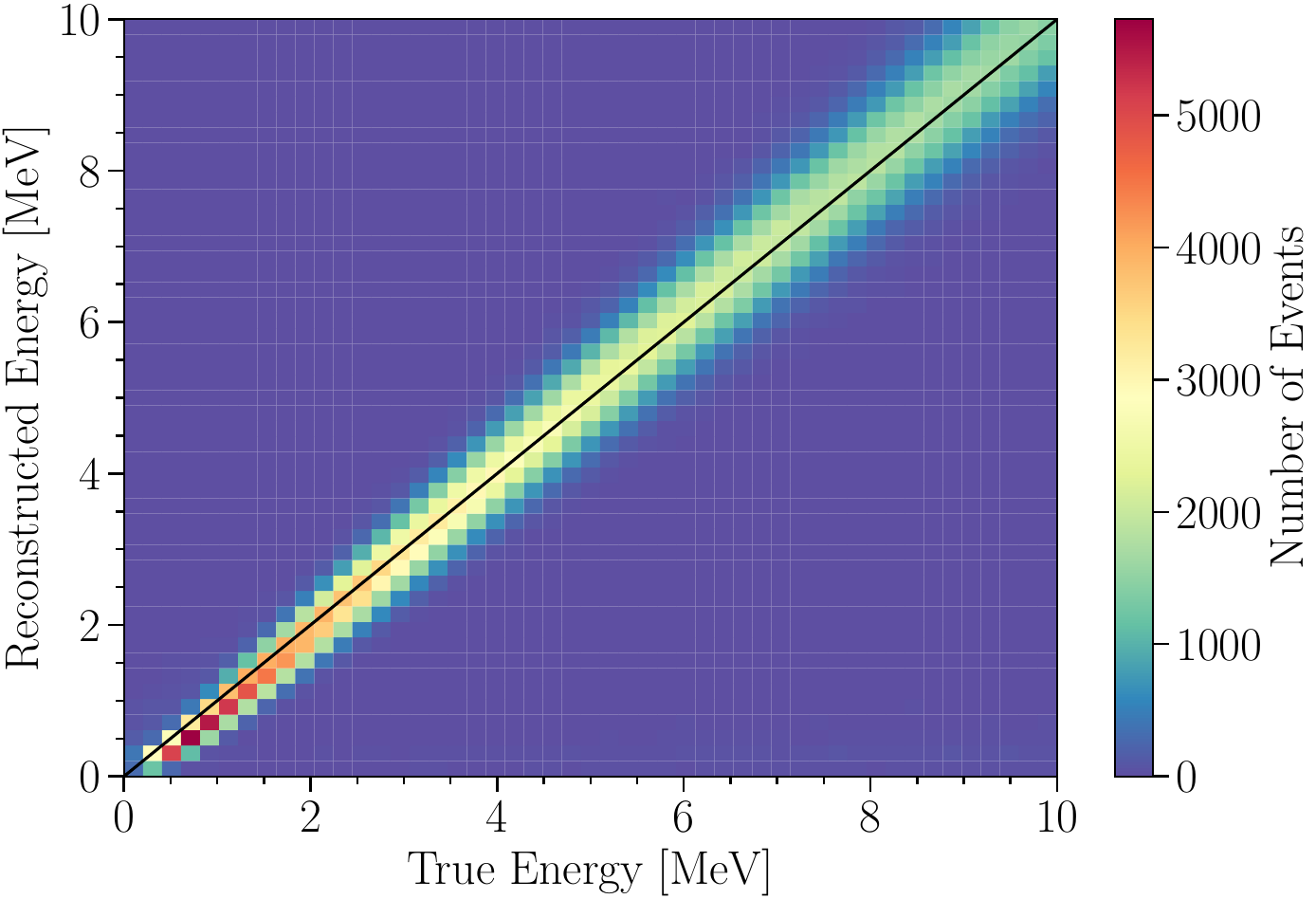}
    \caption{True versus reconstructed energies for simulated events. The event position is first reconstructed with the \texttt{GraphNeT} model, and the energy is then estimated from the charge recorded in the first 90~ns using the position-dependent calibration factors of Fig.~\ref{fig:spline_scaling_vs_position}.}
    \label{fig:true_vs_reco_energy}
\end{figure}

\begin{figure}[h]
    \centering
    \includegraphics[width=\linewidth]{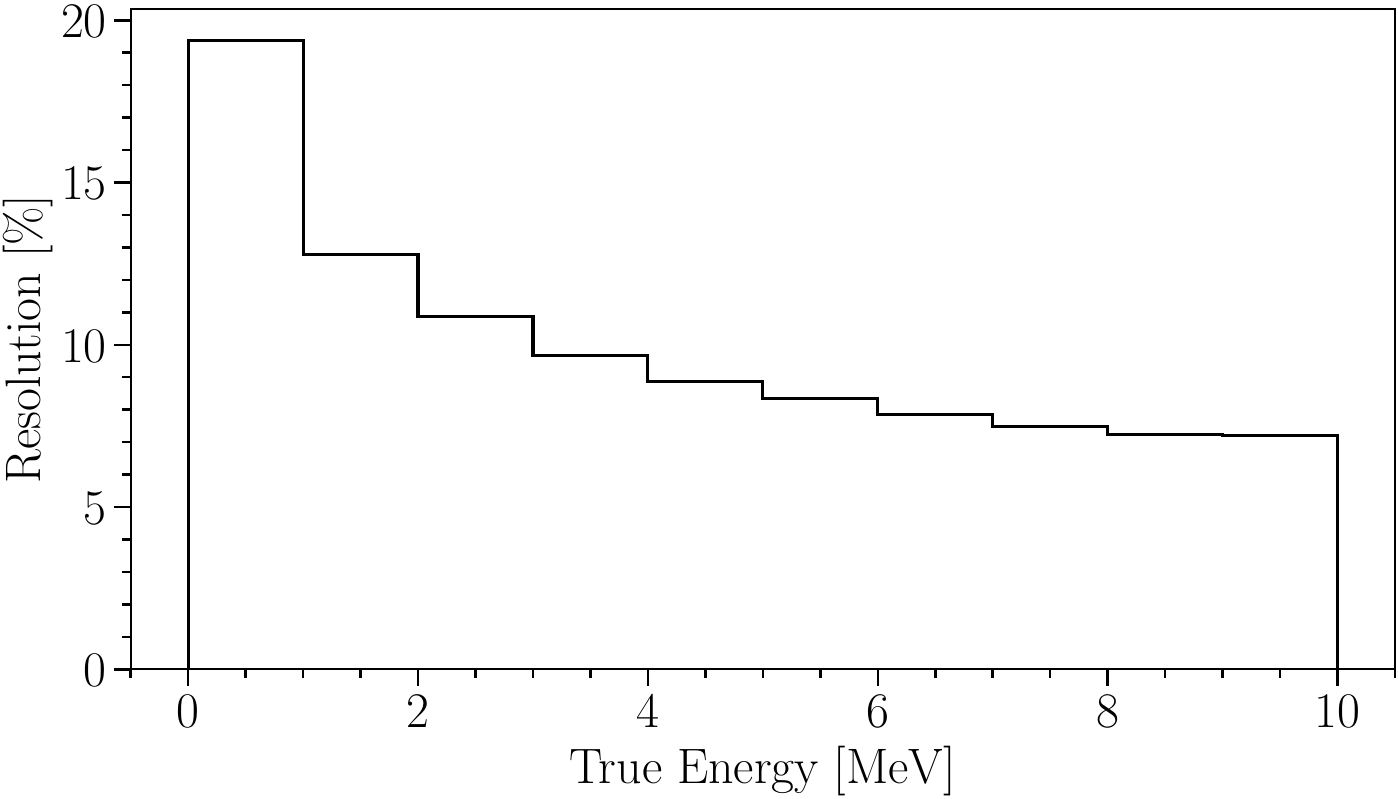}
    \caption{Energy reconstruction resolution versus true energy. The resolution is evaluated using only particles with true kinetic energies $\leq 10$~MeV, corresponding to the energy range relevant for the analysis.}
    \label{fig:resolution_energy}
\end{figure}

The energy reconstruction is evaluated using simulated electrons with true energies below 10~MeV. Fig.~\ref{fig:true_vs_reco_energy} compares the reconstructed and true energies after applying the standard fiducial selection. A strong linear correlation demonstrates that the position-dependent calibration provides an unbiased estimate of the deposited energy across the energy range of interest.

\begin{figure}[h]
    \centering
    \includegraphics[width=\linewidth]{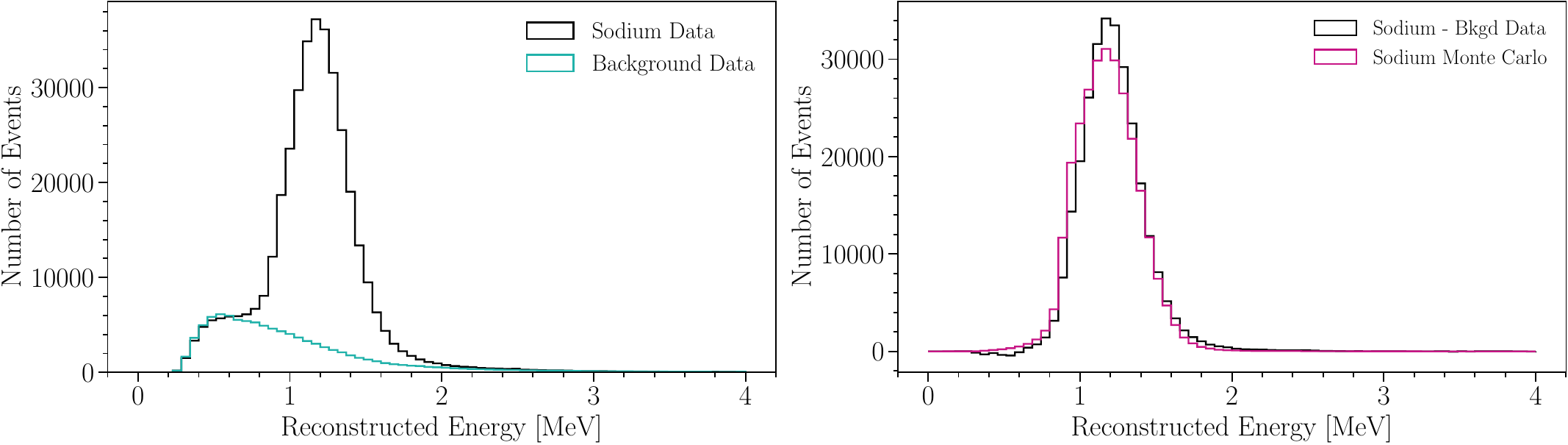}
    \caption{Energy reconstruction validation with the $^{22}$Na calibration source. The left panel compares the reconstructed energy distributions for calibration and background data, while the right panel shows the background-subtracted calibration spectrum overlaid with the Monte Carlo simulation. The reconstructed energy peaks at approximately 1.2 MeV in both data and simulation, with spectral shape differences of less than 10\%.}
    \label{fig:sodium_datamc_energy}
\end{figure}

The corresponding energy resolution is shown in Fig.~\ref{fig:resolution_energy}, where it is defined as the width of the central 68\% interval of $(E_{\mathrm{reco}}-E_{\mathrm{true}})$ divided by the true energy. The resolution improves from approximately 12.5\% at 1~MeV to 7.5\% at 10~MeV, consistent with the expected stochastic scaling of scintillation detectors. This performance provides a reliable energy estimator for the low-energy events considered in the ALP search.

The energy reconstruction is validated using the $^{22}$Na calibration dataset. Fig.~\ref{fig:sodium_datamc_energy} compares the reconstructed energy spectra for source and background data (left), showing a clear excess from $^{22}$Na decays. After subtracting the normalized background, the resulting spectrum is compared with the Monte Carlo prediction, as shown in the right panel of Fig.~\ref{fig:sodium_datamc_energy}.

The reconstructed energy spectra in data and simulation exhibit good agreement in both the energy scale and spectral shape. The peak positions are consistent at approximately 1.2~MeV, while the spectral shapes agree to within $\sim$10\%, consistent with the optical model uncertainties discussed in Ref.~\cite{CCM:2025dbq}. This agreement demonstrates that the position-dependent energy calibration accurately models the detector response and provides a reliable energy estimator for kinematic ranges relevant to this analysis.

\bibliographystyle{apsrev4-2}
\bibliography{main}

\end{document}